\documentclass[aps,pre,twocolumn,amsmath,amsfonts,floatfix,superscriptaddress]{revtex4-1}
\usepackage{natbib}
\usepackage{graphicx,amssymb}
\usepackage{color}

\newcommand{\rev}[1]{\textcolor{black}{#1}}

\begin{document}

\title{How to \rev{``measure''} a structural relaxation time that is too long to be measured?}

\author{L. Berthier}

\email{ludovic.berthier@umontpellier.fr}

\affiliation{Laboratoire Charles Coulomb (L2C), Universit\'e de Montpellier, CNRS, 34095 Montpellier, France.}

\affiliation{Department of Chemistry, University of Cambridge, Lensfield Road, Cambridge CB2 1EW, United Kingdom.}

\author{M. D. Ediger}

\affiliation{Department of Chemistry, University of Wisconsin-Madison, Madison, Wisconsin 53706, USA}

\begin{abstract}
It has recently become possible to prepare ultrastable glassy materials characterised by structural relaxation times which vastly exceed the duration of any feasible experiment. Similarly, new algorithms have led to the production of ultrastable computer glasses. Is it possible to obtain a reliable estimate of a structural relaxation time that is too long to be measured? We review, organise, and critically discuss various methods to estimate very long relaxation times. We also perform computer simulations of three dimensional ultrastable hard spheres glasses to test and quantitatively compare some of these methods for a single model system. \rev{The various estimation methods disagree significantly and it is not yet clear how to accurately estimate extremely long relaxation times.}
\end{abstract}

\date{\today}

\maketitle

\section{The problem of long timescales}

Glassy materials are characterized by long relaxation times~\cite{ediger1996supercooled,berthier2016glass,mckenna2020glass}.  The microscopic origin of these long relaxation times has been a subject of intense interest.  Current approaches suggest that  the relaxation time might be controlled by the configurational entropy~\cite{adam1965temperature,KTW89}, or by dynamical facilitation~\cite{garrahan2002geometrical}, or by the shear modulus~\cite{gundermann2014dynamic,hecksher2015review}, or by one of several additional factors. For most glassformers, the relaxation time of the supercooled liquid state increases in a super-Arrhenius manner as the temperature decreases.  Some experiments and simulations have been interpreted to indicate that the relaxation time would diverge well above absolute zero if the supercooled liquid could be cooled sufficiently slowly, but this conclusion has also been contested~\cite{ediger1996supercooled}.  Aside from the issue of a true divergence, the structural relaxation time of an amorphous system is a key factor determining its material properties. It is a long structural relaxation time that makes an amorphous system have the mechanical properties of a solid. The relaxation time is a critical piece of information needed to understand the barriers on the potential energy landscape and the types of molecular motion responsible for relaxation. 

Conventionally, glasses are produced by cooling bulk liquids below the glass transition temperature, $T_g$, which is the temperature scale set by the competition between the experimental preparation timescale, $t_{prep}$, and the intrinsic equilibrium relaxation time of the system, $\tau_\alpha(T)$. As a general rule, for this method of glass preparation one has 
\begin{equation}
\tau_\alpha(T_g) \approx t_{prep},
\label{eq:coupled}
\end{equation} 
up to a prefactor. In practice, it is difficult to vary $t_{prep}$ over many orders of magnitude. In experiments, $t_{prep} \approx 10^2$~s represents a standard figure. It is possible to realise fast quenches with $t_{prep} = 100$~ms or even shorter times, and a few heroic aging experiments have been performed over months or even years ($t_{prep} = 3 \times 10^7$~s). Computer simulations are limited to about $t_{prep} \leq 1-10 ~\mu$s. 

For liquid-cooled glasses, the preparation time and the relaxation time of the system are strongly coupled, as in Eq.~(\ref{eq:coupled}). It is therefore possible to measure the relaxation time of any such sample, as it takes about the same amount of time to equilibrate the system and to measure its relaxation time. For the past three decades, the experimental challenge has been to devise techniques to measure relaxation times over a broad range of timescales~\cite{dixon1990scaling,adichtchev2003reexamination,blochowicz2003susceptibility} from $\tau_\alpha \approx 100$~s at $T_g$ down to the microscopic relaxation time of simple liquids,  $\tau_\alpha \approx 10^{-10}$~s (faster timescales correspond to the non-glassy fluid). Experiments can now both follow $\tau_\alpha$ over these 12 orders of magnitude, and measure the relaxation spectra of liquids near $T_g$ over a correspondingly large range of frequencies~\cite{lunkenheimer2000glassy,blochowicz2006evolution}.  
  
The paradigm of Eq.~(\ref{eq:coupled}) has recently been shattered as progress in experimental~\cite{swallen2007organic,ediger2017perspective} and numerical~\cite{berthier2016equilibrium,ninarello2017models} techniques has  allowed the rapid preparation of glassy materials with very long relaxation times. For these ``ultrastable'' materials, therefore, the preparation time $t_{prep}$ is typically much shorter than the intrinsic equilibrium relaxation time,
\begin{equation}
t_{prep} \ll \tau_\alpha(T) .
\label{eq:decoupled} 
\end{equation}
In that case, the temperature scale $T_g$ is no longer relevant for the preparation of glasses or for the equilibration of the system. However, $T_g$ remains relevant when it comes to the determination of the relaxation time, because $\tau_\alpha$ can now be much larger than the available measurement time.

\rev{It is of course possible to characterise many physical properties of ultrastable glasses (e.g., density, shear modulus), but the question we would like to ask here is: How can one estimate the (long) relaxation time of these (rapidly prepared) stable glassy systems? Success in estimating these very long relaxation times would enhance our understanding of amorphous materials, allowing more definitive tests of theoretical descriptions~\cite{royall2018race}. Many papers in past decades have attempted to test models and discern fundamental features of supercooled liquids by extrapolating their relaxation times to very low temperatures. We expect that this should be even more successful if the much longer relaxation times of stable glasses could be utilized. Indeed, examples of such efforts have already appeared in the literature~\cite{pogna2015probing,ediger2017perspective,yoon2018testing,ozawa2019does}.} 

Another instance where the problem of long timescales arises is in ``natural aging'' experiments that utilize samples prepared by nature over millions of years. Recently, experimental studies have been performed on amber glasses that were produced naturally about 20-100 million years ago~\cite{love1991influence,zhao2013using,perez2014two,pogna2015probing,pogna2019tracking}, and preserved at ambient temperature since then. For some amber samples, the ambient temperature corresponds to a reasonable fraction of $T_g$ and those glasses have thus aged over about $t_{prep} = 10^{14}$~s, and for them too the available measurement time is much shorter than the relaxation time as indicated in Eq.~(\ref{eq:decoupled}). 

In this article, we restrict our attention to very stable amorphous systems that are equilibrated into the supercooled liquid at some temperature $T<T_g$. This is not the case for some of the experimental stable glasses discussed above. Nevertheless, our results could be applied to such systems through the use of a fictive temperature~\cite{tool1946relation} to map non-equilibrium systems onto an equivalent equilibrium system at temperature $T_f$, before asking what is the relaxation time of an equilibrium system at fictive temperature $T_f$. Therefore, we do not discuss this issue further here, and consider the simpler problem of an equilibrium system characterised by a very long relaxation time. 

Although we discuss in parallel simulations and experiments, the challenge is somewhat different in these two areas. Simulations can now provide equilibrated samples at temperatures below the experimental glass temperature~\cite{berthier2017configurational}, just as experiments do. However, the dynamic window for simulations is $\tau_\alpha \lesssim 10~\mu$s, whereas it is for $\tau_\alpha \lesssim 100$~s for experiments.
In simulations, we thus attempt to estimate relaxation times from $10~\mu$s to $100$~s, whereas for experiments we extrapolate beyond $10^2$~s to estimate relaxation times up to $10^{10}$~s. In both cases, we thus attempt to extend direct measurements by about 7-10 orders of magnitude, but the temperature regimes where the extrapolations are performed are different. 

Various methods to estimate very long relaxation times have been proposed and used in different types of glassy materials. Our first goal is to review and organise these different approaches. For each method, we discuss the practical advantages and inconveniences, and we also critically assess the physical content and underlying hypothesis needed to obtain a measurement. In addition, we also perform an extensive set of computer simulations using a single model system of polydisperse hard spheres in $d=3$ dimensions to compare several methods in a single system, which is typically impossible experimentally.

The numerical model utilized in our simulations is described in Sec.~\ref{sec:model}.
Since all methods necessarily require some kind of extrapolation, we organise the manuscript depending on whether the extrapolation is performed using fully equilibrium conditions (Sec.~\ref{sec:equilibrium}), or not (Sec.~\ref{sec:nonequilibrium}). We summarise our results in Sec.~\ref{sec:conclusion}.

\section{Numerical model}

\label{sec:model}

We study the canonical model of hard spheres in $d=3$
dimensions. The pair interaction is zero for nonoverlapping
particles and infinite otherwise. We use a continuous size
polydispersity~\cite{berthier2016equilibrium}, with a flat distribution of diameters between $\sigma_m =0.63$ and $\sigma_M=1.4$, and a non-additive interaction to avoid crystallisation~\cite{ninarello2017models}. The unit length corresponds therefore roughly to the center of the particle size distribution. We mostly simulate systems composed of $N=10^3$ particles, in a cubic cell with periodic boundary conditions. 

We perform Monte Carlo simulations at constant applied pressure $P$, and record the volume fraction $\phi$. We can alternatively and equivalently perform constant volume simulations at volume fraction $\phi$, and measure the pressure $P$. For hard spheres, pressure and temperature are not independent parameters, and the parameter space is effectively one-dimensional. One can choose to represent the evolution of the system as a function of volume fraction, $\phi$, as is done in most colloidal experiments. Equivalently, the evolution can be represented as a function of the reduced pressure~\cite{berthier2009glass}, 
\begin{equation}
Z = \frac{P}{\rho k_B T},
\label{eq:Z}
\end{equation}
where $\rho = N / V$ is the number density, and $k_B T$ is the thermal energy. We set $k_B=1$. The volume fraction $\phi$ and the reduced pressure $Z$ are one-to-one related by the equilibrium equation of state, $Z=Z(\phi)$. The distinction between isobaric and isochoric paths is thus \rev{immaterial in Monte Carlo simulations~\cite{berthier2009glass}.}

As demonstrated by Eq.~(\ref{eq:Z}), it is equivalent to think of hard sphere simulations as being performed at an imposed pressure $P$, with $T$ arbitrarily set to unity, or at an imposed temperature $T$, with pressure $P$ arbitrarily set to unity, as the physics is only controlled by their ratio. In the latter case, it is obvious that $Z$ is nothing but the inverse temperature, $Z \sim 1/T$. In the following, we follow the evolution of the equilibrium relaxation time $\tau_\alpha$ as a function of $Z$, and this is equivalent to showing $\tau_\alpha$ versus $1/T$ in a fluid with soft interactions~\cite{berthier2009glass}. \rev{When this convention is used, hard spheres and simple continuous potentials such as soft spheres or Lennard-Jones particles appear to be physically very similar~\cite{berthier2009glass,isobe2016applicability,ozawa2019does,berthier2019configurational}. Therefore hard spheres represent a convenient model of a glass-former which can be efficiently studied numerically.}

Times are reported using standard Monte Carlo time steps, where one step represents $N$ attempts to make an elementary translational move. In each elementary step, a particle is chosen at random, and a random displacement $\delta {\bf r}$ of maximal amplitude $| \delta {\bf r} | < 0.05$ is proposed, and accepted if it does not create an overlap with a neighboring particle. \rev{For constant pressure simulations, we perform one attempt to change the linear size of the box every $N$ single particle moves.}

In such ordinary Monte Carlo simulations, the relaxation time $\tau_\alpha(Z)$ of the system represents the structural relaxation of the material~\cite{berthier2007monte,berthier2007revisiting}, as in traditional molecular dynamics simulations. \rev{This equivalence holds provided attention is paid that the acceptance probability of the Monte Carlo moves is not a strong function of the temperature.} For traditional Monte Carlo simulations, therefore, Eq.~(\ref{eq:coupled}) holds, in the sense that it is not possible to prepare systems with relaxation times larger than the total simulated time.
 
To rapidly prepare systems with very long relaxation times, we use the swap Monte Carlo algorithm~\cite{grigera2001fast} which dramatically accelerates the equilibration of dense polydisperse fluids~\cite{berthier2016equilibrium,ninarello2017models}. In the swap algorithm, translational moves are performed in alternance with swap moves where a pair of particles is selected at random and their diameters are exchanged if the swap does not create an overlap. When swap moves are used, one Monte Carlo step represents $N$ attempts to perform a Monte Carlo move  (either translational or swap). 

Once an equilibrium system has been prepared, dynamical relaxation is recorded by measuring a dynamical overlap, 
\begin{equation}
Q(t) = \frac{1}{N} \sum_i \theta(a - |{\bf r}_i(t) - {\bf r}_i(0)|),
\label{eq:overlap}
\end{equation}
where ${\bf r}_i(t)$ is the position of particle $i$ at time $t$, $a=0.2$, and $\theta(x)$ is the Heaviside function. The overlap is a convenient analog of the self-intermediate scattering function. We define the relaxation time of the system as $Q(t=\tau_\alpha)=\exp(-1)$, which is also a good measure of the equilibration time of the system. \rev{As usual, this definition assumes that the stretching of the relaxation is not itself a strong function of temperature.}

\begin{figure}
\includegraphics[width=8.5cm]{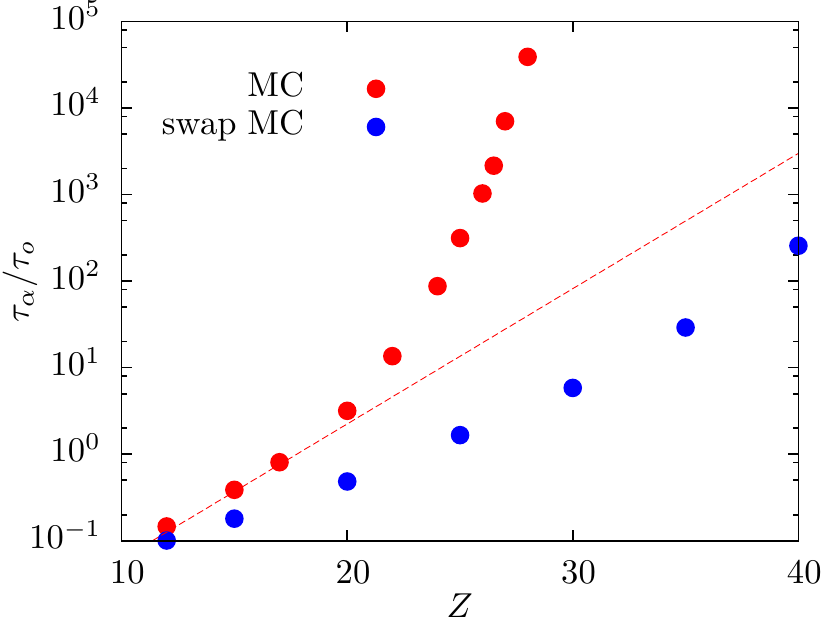}
\caption{Evolution of the equilibrium relaxation time $\tau_\alpha$ of $d=3$ polydisperse hard spheres with reduced pressure $Z \sim 1/T$, using Monte Carlo (MC) simulations. The relaxation time is rescaled by its value at the onset, $\tau_o$, defined by the breakdown of the high temperature Arrhenius fit (dashed line). We use the same $\tau_o$ to rescale the equilibration time measured in the unphysical dynamics of the swap MC, which provides a fast route to prepare equilibrium configurations with large relaxation times.}
\label{fig:bulk}
\end{figure}

In Fig.~\ref{fig:bulk}, we show the equilibrium relaxation time of the simulated system of hard spheres. As with many other model systems, the relaxation time with standard Monte Carlo follows an Arrhenius law at high temperature ($\log \tau_\alpha \sim Z \sim 1/T$ at small $Z$) and becomes super-Arrhenius, or fragile, at low temperatures (large $Z$). It is convenient to use this onset for fragile behaviour to rescale the relaxation time by $\tau_o = 10^4$ Monte Carlo steps, its value at the onset. As can be seen for the red points in Fig.~\ref{fig:bulk}, it is possible to measure $\tau_\alpha/\tau_o$ over about 5 relevant orders of magnitude using conventional numerical simulations. This corresponds to about $10^9$ Monte Carlo steps, and represents about 1 week of CPU time for $N=10^3$ particles.   

Also shown in Fig.~\ref{fig:bulk} is the evolution of the relaxation time when swap Monte Carlo dynamics is used. When the swap moves are present, $\tau_\alpha$ quantifies the very fast equilibration time of the system in a dynamical process that is now distinct from the physical dynamics, and the $\tau_\alpha$ values measured with swap are much shorter. The speedup offered by swap is so important that equilibrating systems at $Z=40$ is extremely fast, whereas conventional simulations cannot equilibrate below $Z \approx 29$. We shall see below that the experimental glass transition corresponding to $\tau_\alpha~\sim 100$~s is near $Z\approx 35$, and swap simulations for this system can reach equilibrium up to $Z \approx 50$, much below $T_g$. Therefore, using swap Monte Carlo, it is possible to prepare very fast (less than $10^6$ Monte Carlo steps, 10 min of CPU time) equilibrium configurations in a regime where the relaxation time is too long to be measured in computer simulations (more than $10^{16}$ Monte Carlo steps, about $10^5$ years of CPU time).
 
In the rest of the article, we will use this hard sphere system to benchmark, illustrate and compare various methods to measure a relaxation time that is too long to be directly measured. We will mainly analyse the regime $Z = 29-40$ which is comfortably explored in equilibrium conditions using the swap Monte Carlo algorithm.

A rough dictionary between numerical and experimental timescales is useful. Typically, for molecular liquids, the onset timescale is near $\tau_o \sim 10^{-10}$~s~\cite{schmidtke2012boiling}. The relaxation time at $T_g$ is 
$\tau_\alpha(T_g) \sim 10^2$~s~$\Leftrightarrow \tau_\alpha / \tau_o \sim 10^{12}$.
This implies that $\tau_\alpha/\tau_o \sim 10^5 \Leftrightarrow \tau_\alpha \sim 10~\mu$s, which is the maximal time we can directly simulate.
In what follows, we frequently extrapolate numerical timescales up to $\tau_\alpha/\tau_o \sim 10^{12}$, which corresponds to $\tau_\alpha = 100$~s, and thus to the experimental glass transition temperature $T_g$.    

\section{Extrapolation using equilibrium conditions}

\label{sec:equilibrium}

In discussing methods for determining very long relaxation times, we start with those that directly analyse the system in the equilibrium conditions in which it has been prepared at temperature $T$. By construction, an experiment can last at most a duration that we denote $t_{exp}$, and we thus consider situations where $t_{exp} \ll \tau_\alpha(T)$. Any measurement can therefore only record a physical quantity defined over a ``short-time'' $t$. By short, we mean $0 \leq t \leq t_{exp} \ll \tau_\alpha(T)$. Depending on the chosen observable, $t$ can be anywhere between $t=0$ (for a purely static quantity) and $t=t_{exp}$, where the physical observable is defined over the maximal duration of the measurement. We organise the corresponding methods by increasing values of $t$ between these two limits, exploring the various possible regimes.  

\subsection{Extrapolation from a static quantity: $t=0$}

\label{sec:temperature}

An obvious choice to extrapolate dynamic data in equilibrium conditions is the temperature, $T$. The method is conceptually very simple. One measures the evolution of $\tau_\alpha(T)$ in the temperature regime where this can be done, i.e. when $\tau_\alpha(T) \leq t_{exp}$. One then fits the relation $\tau_\alpha = \tau_\alpha(T)$ to some appropriate functional form, which is then extrapolated to infer the relaxation time in the inaccessible time regime where $\tau_\alpha(T) > t_{exp}$. 

The question we ask in this subsection is different from the long-standing  debates~\cite{stickel1995dynamics,elmatad2010corresponding,hecksher2008little,mauro2009viscosity,mckenna2015accumulating} regarding the best description of experimental values of $\tau_\alpha$ for the temperature range extending down to $T_g$. While these debates do inform us about possible functional forms, our goal is to find the best fitting functions to extrapolate the dynamic data by some 5-10 orders of magnitude in the most precise manner.

\begin{figure}
\includegraphics[width=8.5cm]{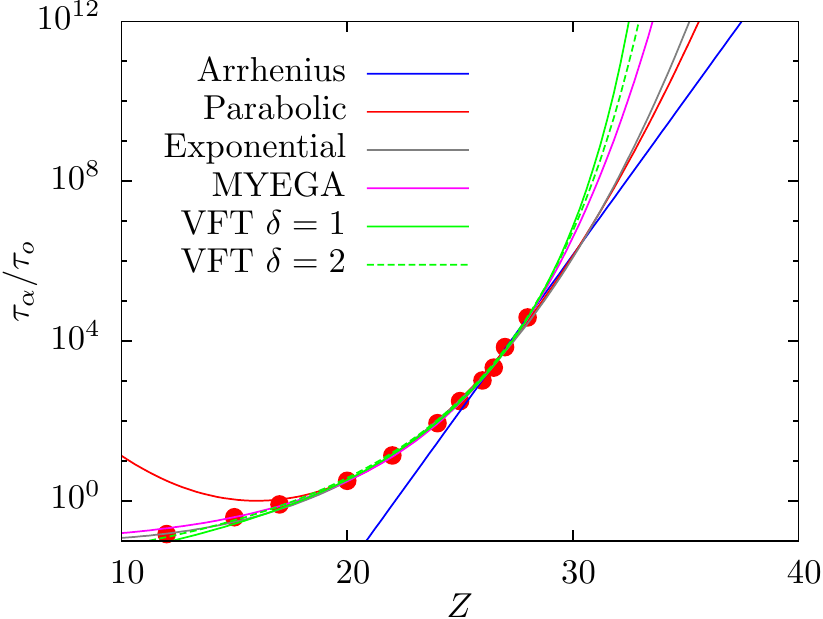}
\caption{Extrapolation of the relaxation time directly measured in hard sphere simulations in the regime $1 \leq \tau_\alpha / \tau_o \leq 10^5$ (symbols) up to the experimental glass transition at $\tau_\alpha/\tau_o = 10^{12}$, using various functional forms presented in the text.}
\label{fig:fit}
\end{figure}

In Fig.~\ref{fig:fit} we present the result of such an analysis for the simulated hard sphere system. We extend the vertical range of timescales from the directly accessible regime, $\tau_\alpha/\tau_o \leq 10^5$, up to the experimental glass transition at $\tau_\alpha/\tau_o = 10^{12}$, and use various functional forms to extrapolate the dynamics down to $T_g$. 

All the functional forms that we use involve activated dynamics, 
\begin{equation}
\tau_\alpha = \tau_\infty\exp \left( \frac{E(T)}{T}\right), 
\label{eq:activation}
\end{equation} 
where $\tau_\infty$ is a constant. The art of fitting then lies in a proper choice for the temperature evolution of the apparent activation energy $E(T)$ in Eq.~(\ref{eq:activation}). We use several such choices. 

The simplest choice is the Arrhenius law, where $E(T)=E=const$. Given that our system shows fragile behavior, the Arrhenius law in Fig.~\ref{fig:fit} can be considered a lower bound to the real growth of the relaxation time. 

The simplest correction to the Arrhenius behaviour would be a second order polynomial for the evolution of $\log \tau_\alpha$ versus $1/T$. This is captured by the parabolic law shown in Fig.~\ref{fig:fit}, $\log \tau_\alpha \sim (Z-Z_o)^2 \sim 1/T^2$. Although introduced and discussed in the context of kinetic facilitation~\cite{elmatad2010corresponding}, fitting to the parabolic law with free parameters is a simple and agnostic way of capturing the fragility of the system. In a similar spirit, the non-Arrhenius behaviour of the dynamics can be captured using a non-analytic function of $1/T$ using the exponential~\cite{avramov1988effect} (denoted `Exponential' in Fig.~\ref{fig:fit}) form $\log\tau_\alpha \sim (E_0/T)^n$ where $n$ is a non-integer exponent (here, we use $n=3.8$). As shown in Fig.~\ref{fig:fit}, this provides an extrapolation close to the parabolic form for the hard sphere system. As we discuss below, we expect that these two functions represent the most faithful extrapolations as they simultaneously capture the fragility of the actual data (unlike the Arrhenius form) and simply extrapolate the measured curvature over the next few decades.

We add two other fitting forms which become strongly divergent at lower temperatures. One is the double exponential fit~\cite{mauro2009viscosity} (denoted `MYEGA' in Fig.~\ref{fig:fit}) where $\log \tau_\alpha \sim Z \exp(aZ)$ which does not introduce a finite temperature singularity, but leads to a very strong increase of $\tau_\alpha$ at very low $T$. A second family of functions (denoted 'VFT' for Vogel-Fulcher-Tamman in Fig.~\ref{fig:fit}) introduces a finite temperature dynamic singularity, $\log \tau_\alpha \sim A / (Z-Z_0)^\delta$, with an exponent $\delta$ which can also vary. We show fits to the MYEGA form and to a dynamic VFT singularity using both $\delta=1$ and $\delta=2$ in Fig.~\ref{fig:fit}. These three functions closely follow each other. The dynamic singularities estimated using the VFT laws occur at $Z_0=38$ (for $\delta=1$) and $45$ (for $\delta=2$). Since the swap algorithm can equilibrate the system at even lower temperatures without ever encountering any phase transition, we must conclude that the extrapolated timescales using the VFT law (and thus, using MYEGA) overestimate the real behaviour of this system. 

Despite the quantitative differences obtained between all types of extrapolations in Fig.~\ref{fig:fit}, we note that the spread in the location of the estimated glass transition temperature $T_g$ defined as $\tau_\alpha/\tau_o=10^{12}$ remains modest, the spread being in the range $Z \in [32,37]$, with the most sensible fitting functions (Exponential and Parabolic) falling  in the middle of that range, near $Z_g \approx 35$. Therefore, we conclude that extrapolating numerical data from 5 to 12 orders of magnitude remains a relatively safe exercise if the goal is to locate the experimental glass transition temperature. A less optimistic view is to consider the spread in extrapolated relaxation times near $Z=33$, which covers about 5 orders of magnitude, between the estimates that we consider as lower and upper bounds. 

In previous work~\cite{ozawa2019does}, we used experimental data to perform the exercise above, using direct measurements in the range $\tau_\alpha/\tau_o=1-10^5$ to extrapolate the dynamic up to $\tau_\alpha/\tau_o=10^{12}$, where experimental measurements can still be done. We found that the parabolic extrapolation performed very well indeed~\cite{elmatad2010corresponding}. In the rest of the paper we choose the parabolic fit as a sensible extrapolation and use it to provide a reference against which we compare the other extrapolations, but our conclusions do not depend on this specific choice. 

\begin{figure}
\includegraphics[width=8.5cm]{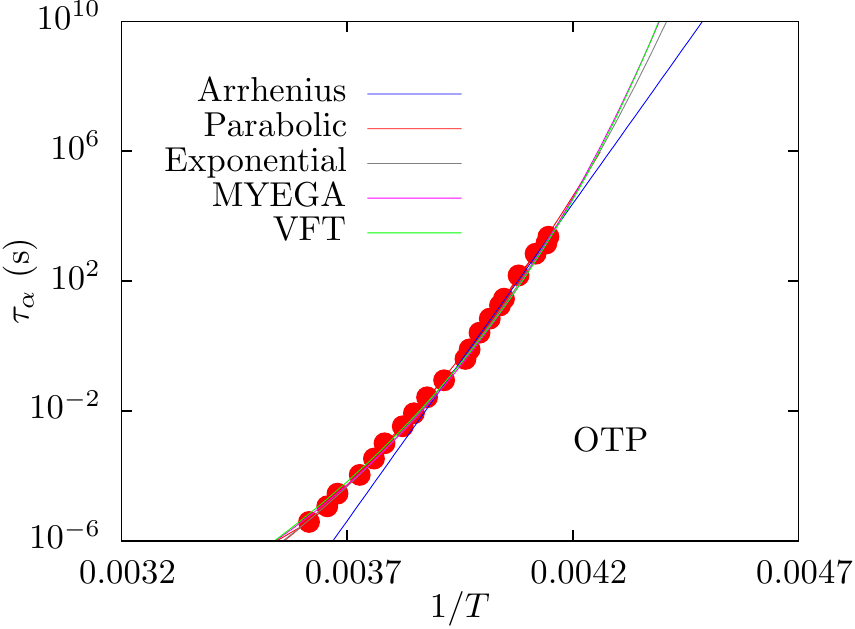}
\includegraphics[width=8.5cm]{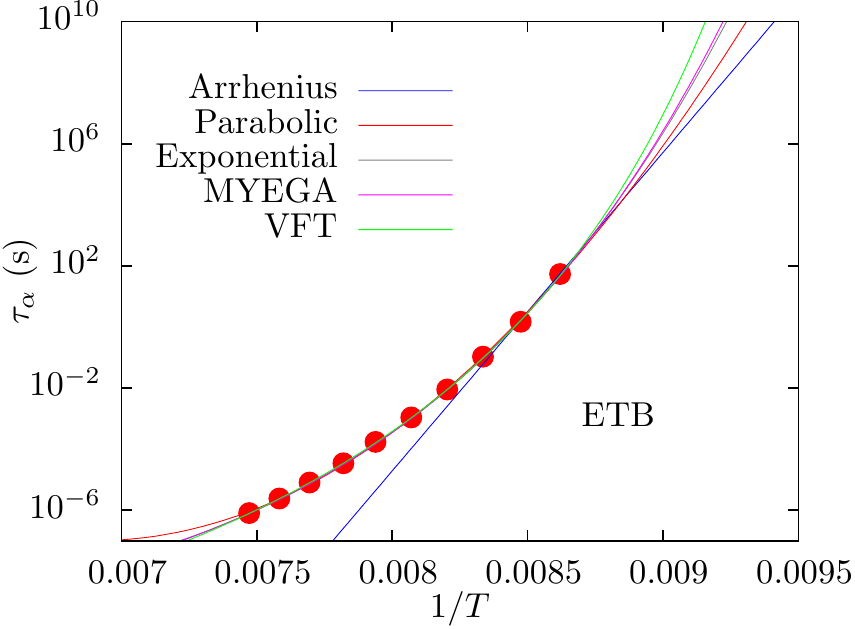}
\caption{Extrapolation of the relaxation time directly measured in experiments
on orthoterphenyl (OTP~\cite{hansen1997dynamics}) and ethylbenzene (ETB~\cite{chen2011dynamics}) in the regime $10^{-6} \leq \tau_\alpha \leq 10^2$~s (symbols) up to the $\tau_\alpha = 10^{10}$~s, using various functional forms discussed in the text.}
\label{fig:otp}
\end{figure}

We can follow the same extrapolation recipe with experiments, using actual measurements in the regime $\tau_\alpha = 10^{-6}-10^{2}$~s to estimate the dynamics up to $\tau_\alpha = 10^{10}$~s~$\approx 300$~years. We report the result of this exercise for two materials in Fig.~\ref{fig:otp}: orthoterphenyl (OTP~\cite{hansen1997dynamics}) and ethylbenzene (ETB~\cite{chen2011dynamics}), using the same functional forms (Arrhenius, Exponential, MYEGA, VFT). As with the simulations, we find that the spread between the various fitting functions is rather modest for an extrapolation of 8 orders of magnitude, with Arrhenius and VFT again appearing at the boundaries. Clearly, the spread is more modest for OTP than it is for ETB, which presumably stems from the different curvature of the data measured above $T_g$ which is quite modest for OTP and constrains the fits more strongly.

Although we have chosen temperature as the simplest example of a static observable, we note that other quantities could be used as well. Among well-known examples we can list the static structure factor (which can be measured both in simulations and experiments) or the number of locally favoured structures (easily measured in simulations or colloidal experiments). Soft vibrational modes (encoded in the Hessian matrix) and machine learned softness have also been found to correlate well with the dynamics~\cite{widmer2006predicting,widmer2008irreversible,schoenholz2016structural}. Such alternatives could potentially be useful, but one would need to have a solid theoretical prediction (or some empirical fit, or well-tested model for $T>T_g$) describing how the chosen static quantity relates to the relaxation time. In the absence of an accepted predictive theory, one would need strong evidence that some particular static quantity correlates in a universal manner with the relaxation time for every substance. 

Among thermodynamic observables, the configurational entropy $S_{conf}(T)$ plays a special role in glass physics~\cite{berthier2019configurational}. Its strong temperature dependence is viewed as an important empirical signature of supercooled liquids~\cite{Kauz48,ediger1996supercooled}. The configurational entropy plays a central role in some theoretical descriptions~\cite{KTW89,LW07,parisi2020theory}. In addition, since the work of Adam and Gibbs~\cite{adam1965temperature}, the idea that the decrease of configurational entropy can be directly related to the growth of the structural relaxation time has been frequently revisited~\cite{KTW89,bouchaud2004adam} and tested~\cite{richert1998dynamics,sastry2001relationship}. In its original form, the Adam-Gibbs relation states that 
\begin{equation}
\log (\tau_\alpha / \tau_o ) = A / (T S_{conf}(T)). 
\label{eq:AG}
\end{equation}
In the present context, Eq.~(\ref{eq:AG}) potentially allows the determination of $\tau_\alpha$ from the sole determination of a thermodynamic (static) quantity, $S_{conf}(T)$. In computer studies, configurational entropy has recently been measured in several models at very low temperatures~\cite{berthier2017configurational,ozawa2018configurational,berthier2019zero,berthier2019configurational}. However, a recent study dedicated to the Adam-Gibbs relation suggests that Eq.~(\ref{eq:AG}) would lead to incorrect estimates of the relaxation time~\cite{ozawa2019does}. For $d=3$ hard spheres in particular, directly using Eq.~(\ref{eq:AG}) to extrapolate $\tau_\alpha$ strongly overestimates its temperature dependence. 
This idea has not yet been exploited experimentally, because experimental determinations of the configurational entropy in either naturally aged glasses or vapor deposited films are not available, but this issue clearly deserves further study.  

{\it Critical discussion.} The method studied in this subsection is conceptually simple, but relies on the choice of a ``best'' fit to extrapolate the data. Even when a reasonable fit is performed, its validity in the regime where no data exists is of course questionable, and cannot be proven. Implicitly, the method also assumes that the physics at play does not change in the extrapolated regime, and neglects for instance the possibility of a pronounced fragile-to-strong crossover, which can occur if the microscopic mechanism for relaxation evolves qualitatively in the regime where the dynamics needs to be extrapolated. 

There is no consensus at the moment about how to best fit the temperature evolution of $\tau_\alpha$, and several empirical fits work very well.  However, there is a distinction between choosing ``the'' temperature dependence of $\tau_\alpha$ which requires a deep understanding of the physics, and choosing the best functional form to extrapolate the data by some orders of magnitude, which is a more agnostic exercise which only requires the physics to be about right, i.e. thermally activated dynamics with a smooth functional form for $E(T)$. It could be interesting to employ even more agnostic forms of extrapolation, as proposed for instance in Ref.~\cite{chtchelkatchev2019uncertainty}. A major conclusion from Figs.~\ref{fig:fit} and \ref{fig:otp} is that various functional forms lead to a modest spread of the extrapolations, with well-motivated upper and lower bounds.  

\subsection{From a short time quantity, $t \sim \tau_o$}

\label{sec:DW}

The idea here is to replace the direct extrapolation from a purely static quantity by the measurement of a dynamic quantity that involves only a very ``short'' time $t$, i.e., a timescale which does not grow rapidly as the temperature decreases and remains of the order of the onset timescale $\tau_o$. Physically, such a method must rely on the idea that complete information on the physical processes leading to structural relaxation is already encoded in a short time observable.

Several examples of such observables have been discussed in the literature, 
such as the Debye-Waller factor~\cite{hall1987aperiodic} (amplitude of short-time vibrational motion), the non-ergocity factor~\cite{scopigno2003fragility} (short-time plateau in density-density correlations), the shear modulus~\cite{dyre2006colloquium} (plateau value of stress-stress autocorrelation function), the density of states~\cite{zaccone2020relaxation} (distribution of eigenvalues of vibrational modes), and the speed of sound~\cite{pogna2015probing} (measured in scattering experiments). All these quantities are directly related to short-time motion occurring at the particle scale in glasses at low temperatures, from which long-time structural relaxation is inferred. These observables typically refer to dynamics occurring on a timescale of about 1~ns over which very little (if any) of the structural relaxation has occurred at very low temperatures. 

If such a short time quantity could be used to estimate very long relaxation times, this would of course be very useful, as a very long measurement could be replaced by a very short one. This would of course solve the problem posed by Eq.~(\ref{eq:decoupled}). For this approach to work, however, one needs to precisely know the connection between the short-time quantity and the relaxation time. The discussion of possible connections of this type is a long-debated area in the field of the glass transition~\cite{dyre2006colloquium}, and this line of research navigates between empirical discoveries, phenomenological models, and physical arguments. The shoving model~\cite{dyre2006colloquium} is an example of an approach that uses the shear modulus to predict the structural relaxation time, and was tested experimentally in several studies~\cite{maggi2008supercooled,gundermann2014dynamic,hansen2017connection}. Leporini {\it et al.} have provided an empirical fit between the DW factor and the relaxation time~\cite{larini2008universal}, which extends a relation proposed previously for network glasses~\cite{hall1987aperiodic}. An empirical modification of this model was proposed in Ref.~\cite{simmons2012generalized}.
 
\begin{figure}
\includegraphics[width=4.25cm]{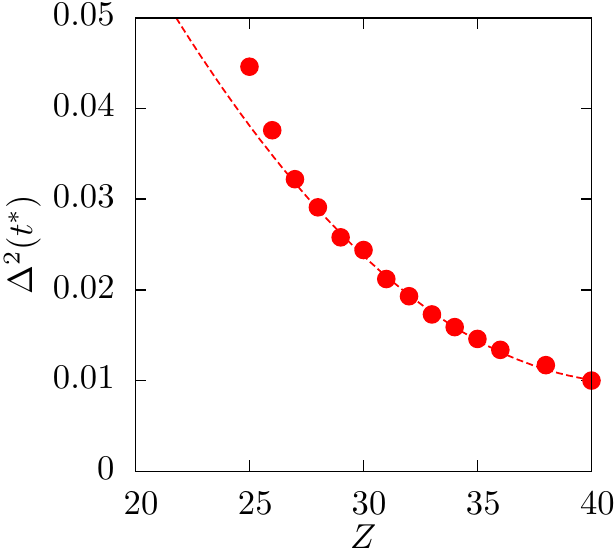}
\includegraphics[width=4.25cm]{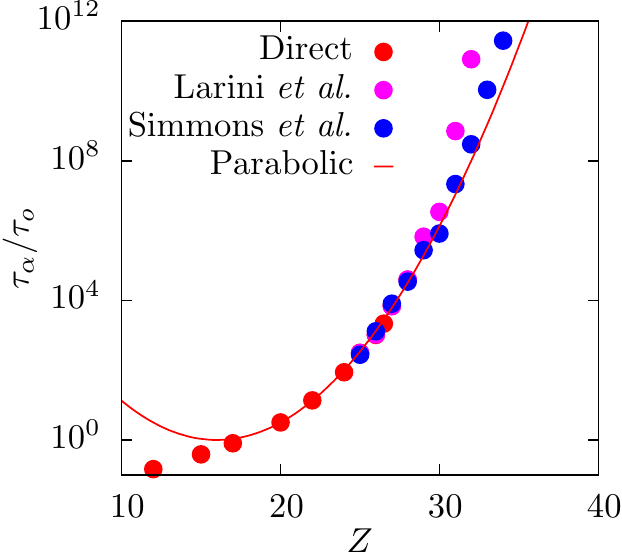}
\caption{Extrapolation of the relaxation time in hard sphere simulations using the Debye Waller (DW) factor. Left: smooth evolution of the DW factor with inverse temperature, $Z \sim 1/T$.  
Right: the parametric relation $\tau_\alpha(DW)$ is fitted using two functional forms drawn from Refs.~\cite{simmons2012generalized,larini2008universal}
in the numerical regime, and extrapolated to low temperatures. The extrapolations compare reasonably well with the parabolic fit.}
\label{fig:lepo}
\end{figure}

Numerically, the easiest way to apply such a method is to measure the short-time plateau value of the mean-squared displacement in equilibrium conditions, because this observable requires the least computational effort. The mean-squared displacement is defined as 
\begin{equation}
\Delta^2(t) = \frac{1}{N} \sum_i \langle |{\bf r}_i(t) -{\bf r}_i(0)|^2 \rangle.
\end{equation}
The Debye-Waller (DW) factor is conventionally defined as the value of $\Delta^2(t)$ at a time $t^*$ where the MSD displays an inflection point, i.e. when $\partial \log \Delta^2(t)/ \partial \log t$ exhibits a minimum~\cite{larini2008universal}. Using the dictionary above to connect the numerical and experimental timescales, this inflection point occurs roughly in the regime $10^4 - 10^5$ Monte Carlo time steps, which indeed corresponds to about $0.1-1$~ns. 
 
The evolution of $\langle \Delta^2(t^*)\rangle$ with pressure for the hard sphere system is shown in Fig.~\ref{fig:lepo}, which demonstrates that the DW factor decreases smoothly as temperature decreases. Previously published simulation results typically cover a much narrower temperature range~\cite{larini2008universal}. We find that neither the $1/T$ nor the $T$ dependence of the DW factor is linear, but the temperature evolution can easily be fitted by second order polynomials using either variables, see Fig.~\ref{fig:lepo}. We see no physical reason to introduce fitting functions containing singularities in the temperature dependence of the DW factor (as proposed for instance in Ref.~\cite{simmons2012generalized}).

In the absence of an agreed model to relate the DW factor to the relaxation time, we take a pragmatic approach to the treatment of the data. First, we make a parametric plot of the evolution of $y=\log \tau_\alpha$ versus the inverse of the DW factor, $x = 1/\langle \Delta^2\rangle$ in the accessible numerical regime. We fit the obtained curve  either with a second-order polynomial ($y= a + b x + cx^2$, as in Ref.~\cite{larini2008universal}), or a non-analytic form ($y= a+b x^n$, where $n \approx 1.6$, close to values reported in Ref.~\cite{simmons2012generalized}). Notice that these two fitting functions thus lead to expressions for the relaxation time that are mathematically equivalent to the parabolic and exponential fits used in Sec.~\ref{sec:temperature}. We then use these functions to estimate the relaxation time in a regime where only the DW factor can be measured, see Fig.~\ref{fig:lepo}. By construction, the prediction is excellent in the numerical regime where the relaxation time can be directly measured, and we find that it provides an extrapolation at low temperatures which is somewhat above the extrapolation obtained using the parabolic fit. In particular, the model proposed in Ref.~\cite{larini2008universal} leads to a sligthly faster increase of the relaxation time than the one of Ref.~\cite{simmons2012generalized}, at least for our system. 

Experimentally, an approach using short time dynamics to estimate long relaxation times was followed in Ref.~\cite{pogna2015probing} to analyse vapor-deposited ultrastable glasses. Here, an empirical relation between the derivative of the relaxation time and the speed of sound~\cite{scopigno2003fragility} was used to reconstruct the temperature dependence of the relaxation time in the regime where only the speed of sound can be measured. This allowed the authors to estimate relaxation times up to $10^{10}$~s from speed of sound measurements. 
 
{\it Critical discussion.} The method discussed here exploits the generic idea that the long-time structural relaxation in supercooled liquids is somehow encoded in short-time dynamics measurements. Although sometimes described as a ``growing consensus''~\cite{pogna2015probing}, this idea remains vividly debated. Even if one accepts the existence of correlations, much remains to be done to transform such observations into causal, predictive tools for the dynamics. This implies that the method followed here is essentially empirical and heavily relies on fitting. In our view, there is in fact not much of a conceptual difference between extrapolating a direct fit to the temperature dependence of the relaxation time, or extrapolating an indirect fit of its parametric dependence against, say, the Debye Waller factor. The two approaches are in fact mathematically equivalent. This method would be superior if there were strong evidence for a single, universal function that connects short time dynamics to long relaxation times. 
%ARE YOU OK WITH THIS SENTENCE? IN ESSENCE, WE ARE SAYING THAT THESE TWO GROUPS DID NOT PROVIDE THIS STRONG EVIDENCE - LET'S DISCUSS BRIEFLY.

Our conclusion is that the relaxation time is not better estimated using an extrapolation involving short-time dynamic functions rather than with temperature extrapolations. In particular, claims based on such extrapolations regarding the fate of supercooled liquids at low temperatures are \rev{quite speculative~\cite{pogna2015probing}.} More fundamentally, the theoretical basis for a deep, quantitative connection between short time dynamics and structural relaxation remains to be developed further. 

\subsection{From an intermediate time quantity, $\tau_o \ll t \ll \tau_\alpha$}

\label{sec:beta}

Here, we explore the idea of a measurement performed over a time $t$ that grows as temperature decreases, but not as fast as $\tau_\alpha$ itself, so that such a measurement can still be performed in a regime where $\tau_\alpha$ can no longer be accessed. The method is conceptually not very different from a very short time quantity discussed above, as one still needs to connect the outcome of the measurement over a time $t$ to infer a very large value of $\tau_\alpha \gg t$. 

As a specific example we discuss the $\beta$ process, which is visible at a timescale $\tau_o \ll \tau_\beta(T) \ll \tau_\alpha(T)$. This refers to dynamical processes which do slow down with decreasing temperature, but this slowing down is typically much less pronounced than the one of $\tau_\alpha$. More generally, we may ask whether dynamics occurring over a timescale 
$\tau_\beta  \ll \tau_\alpha$ can be used to infer the behaviour of $\tau_\alpha$, independently of its precise nature or classification (whether this is a $\beta$ process, an excess wing, a secondary process, etc.). By studying the connection $\tau_\alpha(\tau_\beta)$ at equilibrium, one could potentially infer $\tau_\alpha$ by simply measuring $\tau_\beta$.

At the theoretical level, a connection between the timescales $\tau_\beta$ and $\tau_\alpha$ is at the core of the coupling model~\cite{ngai1998relation}, which predicts a relation between three timescales:
\begin{equation}
\tau_\alpha = ( \tau_o^n \tau_\beta)^{1/(1-n)},
\label{eq:coupling}
\end{equation}
where $n$ is a parameter of the model. For $1/(1-n)>1$, one has $\tau_\beta \ll \tau_\alpha$. Clearly, if the model in Eq.~(\ref{eq:coupling}) is well obeyed with a parameter $n$ that is known, then this relation can be directly used to deduce (a very long) $\tau_\alpha$ from the measurement of (an accessible) $\tau_\beta$. 

As an application, Roland and Casilini~\cite{casalini2009aging} have explored this idea in the non-equilibrium glassy state of polyvinylethylene. They measured $\tau_\beta$ in the glassy state below $T_g$ for an ordinary cooling procedure, and inferred $\tau_\alpha$ using an expression close to Eq.~(\ref{eq:coupling}). 
 
To our knowledge, this method has not yet been used to study glasses with large relaxation times such as vapor deposited glasses, although the dynamics at timescales corresponding the $\beta$ relaxation has been analysed~\cite{yu2015suppression,kasting2019relationship}. Importantly, there are many indications that the structural relaxation times in ultrastable glasses may be more than 8 orders of magnitude larger than in ordinary glasses, but the timescale of the $\beta$ process itself is changed by a much smaller factor of at most one order of magnitude~\cite{rodriguez2018distinguishing,ngai2018change}. In contrast, Eq.~(\ref{eq:coupling}) with a reasonable value of $n$ predicts a much larger shift in $\tau_\beta$. A possible explanation for this behaviour was offered~\cite{ngai2018change}. 

{\it Critical discussion.} Unlike static and short time quantities, this type of measurement deals with the dynamics at times that are not microscopic, and as such more directly connects to the slow relaxation processes. However, the empirical relation $\tau_\alpha=\tau_\alpha(\tau_\beta)$ still needs to be extrapolated into unknown territory, and the theoretical basis for this connection provided by the coupling model is not widely accepted. In addition, the timescale $\tau_\beta$ measured experimentally decreases with the aging time for a number of systems Ref.~\cite{casalini2009aging,rodriguez2018distinguishing}, whereas the structural relaxation presumably increases instead, suggesting that care is needed to relate $\tau_\beta$ to $\tau_\alpha$ in an unambiguous manner. All these issues would clearly deserve further experimental analysis.  

\subsection{From the maximal accessible time, $t = t_{exp}$}

\label{sec:TTS}

In this approach, one measures the dynamic relaxation (some correlation or response function, $Q(t)$) over as broad a dynamic range as possible, i.e. up to the maximal time window $t_{exp}$. If the temperature is high enough, then one can directly access the entire relaxation process, from which the relaxation time $\tau_\alpha$ can be estimated, e.g. $Q(t=\tau_\alpha) = \exp(-1)$ for a normalised correlator. When temperature becomes smaller, only a small part of the relaxation process can be observed within the experimental time window. The goal is to then use this small part of the dynamic relaxation process to infer the value of the (inaccessible) $\tau_\alpha$. 

The necessary assumption behind such method is that the relaxation processes are the same at high and low temperatures, up to a global rescaling of the distribution of timescales by a single number, $\tau_\alpha$. This assumption is sometimes called the time temperature superposition (TTS) principle, and has been studied experimentally quite extensively, with some liquids obeying the superposition principle quite well, and some others not at all. This property has also been discussed theoretically in various contexts, but there is no general conclusion since it is sometimes found to hold~\cite{gotze2008complex}, and sometimes found to be violated~\cite{viot2000heterogeneous,xia2001microscopic}. We see no particular reason (from either experiments or theory) to conclude that the superposition hypothesis should be generically obeyed~\cite{niss2018perspective}. 

Mathematically, the time temperature superposition principle assumes that the structural relaxation obeys  
\begin{equation}
Q(t) = {\cal Q}(t/\tau_\alpha),
\label{eq:TTS}
\end{equation}
over a very broad (in log-scale) range of the scaling variable $t/\tau_\alpha$. In practice, the measurement of the scaling function ${\cal Q}(x)$ at high temperature is used to guide the rescaling of the data at low temperatures using $\tau_\alpha$ as a free scaling parameter along the horizontal time axis.

\begin{figure}
\includegraphics[width=4.25cm]{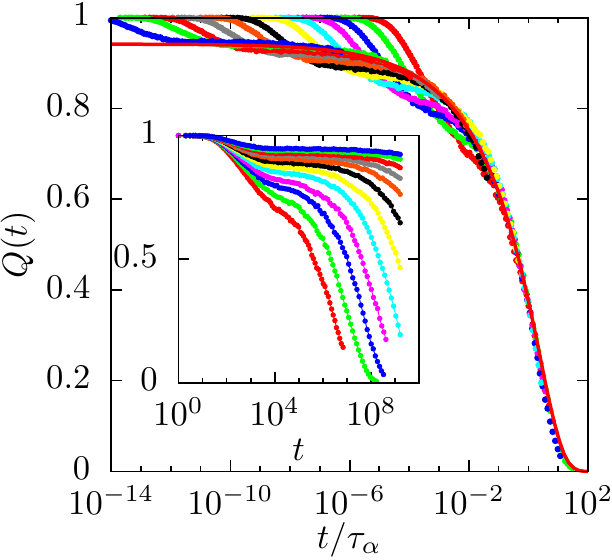}
\includegraphics[width=4.25cm]{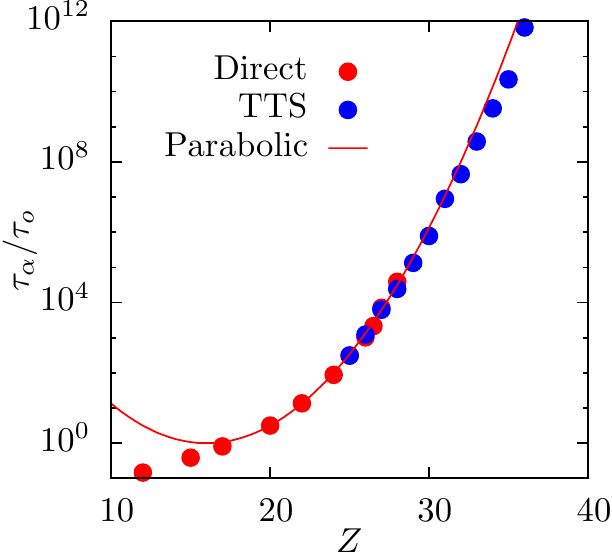}
\caption{Extrapolation of the relaxation time in hard sphere simulations using time temperature superposition (TTS) principle in Eq.~(\ref{eq:TTS}). Left: measured time correlation function $Q(t)$ using either $t$ (inset) and $t/\tau_\alpha$ (main) \rev{from $Z=25$ to $Z=36$ (left to right). The same colors are used in later figures.} The full red line is the mastercurve ${\cal Q}(x)$. Right: comparison of measured and extrapolated $\tau_\alpha$ along with the parabolic fit.}
\label{fig:mc}
\end{figure}

We illustrate how the method works for the simulated hard sphere system. We first show the dynamic correlation $Q(t)$ measured in equilibrium conditions over a fixed, large time window of $2 \times 10^9$ Monte Carlo steps, see Fig.~\ref{fig:mc} (inset). While the entire decorrelation can be followed at high enough temperatures, we can only detect a very long plateau followed by the onset of decorrelation for the lowest temperature shown in the figure. We then rescale the data using $\tau_\alpha$ as a free scaling factor. In practice, we use the measured $\tau_\alpha$ in the high temperature regime, and we progressively rescale lower temperatures one after the other to construct a mastercurve. The result of the exercise is shown in Fig.~\ref{fig:mc}, where we also show an empirical function that describes the mastercurve ${\cal Q}(x)$ in Eq.~(\ref{eq:TTS}), \rev{with a final stretched exponential decay with stretching exponent $0.75$.} The extrapolated relaxation times are compared to the directly measured $\tau_\alpha$ and the reference parabolic fit in Fig.~\ref{fig:mc}. We see that the extrapolation is very smooth and seems to follow quite closely the parabolic fit.

A limitation of this approach is illustrated by our choice of data sets in Fig.~\ref{fig:mc}. Although we can produce equilibrium configurations at even lower temperatures than those shown, we do not utilize them here because we only detect a very long horizontal plateau in $Q(t)$, and the horizontal shift on the mastercurve becomes too imprecise. For example, for $Z=40$, over the last 5 decades of time, the correlation function decays by a factor of only $\Delta Q / Q \approx 0.6 \%$. 

{\it Critical discussion.} A positive point is that this method exploits the entire dynamic information available to an experiment or a simulation regarding the relaxation process, which takes place entirely in equilibrium. However, this method relies on the unproved assumption that Eq.~(\ref{eq:TTS}) works at all temperatures. This assumption is presumably correct over a narrow enough temperature range, as the physics evolves smoothly with temperature. However, when several orders of magnitude are extrapolated there can be no way to test the validity of the result, which may very well depend on both the considered material and the chosen physical probe to record the dynamics~\cite{echeverria2003enthalpy}. A specific problem is that when the probed dynamics (over a timescale $t_{exp}$) and the extrapolated timescale $\tau_\alpha$ differ by many orders of magnitude, as in Eq.~(\ref{eq:decoupled}), it becomes possible that the dynamics only probes secondary processes (such as excess wings or $\beta$ process) whose temperature dependence is weaker than the one of the structural relaxation time, which would then lead to a strong underestimate of the extrapolated time scale. For ultrastable glasses, it is even possible that some yet unknown secondary process becomes prominent at very low temperatures, and is followed instead of the true structural relaxation.
 
\section{Non-equilibrium and non-linear methods}

\label{sec:nonequilibrium}

In this section, we discuss approaches that use a non-linear perturbing field to speed up the dynamics, so that relaxation times that are too long in equilibrium become measurable due to the influence of the perturbation. These methods also involve an extrapolation of these biased relaxation times to estimate the equilibrium relaxation time in the absence of a non-linear field. The fact that the dynamics speed up markedly indicates that the perturbation is strongly non-linear, and thus the system is indeed relaxing, but these dynamics occur far from equilibrium. 

\subsection{Shear flow}

\label{sec:shear}

When a highly viscous supercooled liquid is subjected to a shear flow, its viscosity remains constant within the linear response regime, but it then decreases rather sharply due to the imposed external flow~\cite{sollich1997rheology,berthier2000two}. This is called shear thinning behaviour~\cite{larson1999structure}. At the microscopic level, relaxation processes also speed up in the non-linear regime, and the relaxation time then becomes a function of the external flow~\cite{yamamoto1998dynamics,berthier2002nonequilibrium}. 

It is easy to observe such behaviour in dense colloidal suspensions and complex fluids, because these systems form soft glasses which can be deformed easily~\cite{crassous2006thermosensitive,pham2008yielding}. The same behaviour is also observed in molecular supercooled liquids and polymeric materials~\cite{lee2009direct,capaldi2002enhanced}, although these materials may eventually break and fracture at large deformations. It is easy to perform computer simulations of the influence of a shear flow in supercooled liquids. 

If the imposed rate of deformation, the shear rate $\dot{\gamma}$, is small enough, then the shear stress $\sigma$ will be given by the Newtonian viscosity $\eta_0(T)$:
\begin{equation}
\sigma = \eta_0(T) \dot{\gamma}.
\label{eq:newton}  
\end{equation} 
Since $\eta_0(T)$ and $\tau_\alpha(T)$ typically have very similar temperature dependences, a rheological measurement is a good way to estimate the relaxation time of the system.
When the shear rate increases, however, the system is driven out of equilibrium. The shear flow then distorts the structure and may accelerate the relaxation of the structure. The measured viscosity is then non-Newtonian, and becomes a function of both the temperature and the imposed shear rate, 
\begin{equation}
\sigma = \eta(T,\dot{\gamma}) \dot{\gamma}, 
\label{eq:shearthinning}
\end{equation}  
where typically $\eta(T,\dot{\gamma}) \sim \tau_\alpha(T,\dot{\gamma}) \sim 1/\dot{\gamma}$. Interestingly, the crossover between linear and non-linear regimes is controlled by $\tau_\alpha(T)$ itself, so that Eq.~(\ref{eq:newton}) is obeyed when $\dot{\gamma} \tau_\alpha (T) \ll 1$, whereas Eq.~(\ref{eq:shearthinning}) holds in the opposite limit, 
$\dot{\gamma} \tau_\alpha(T) \gg 1$. These two limits can be combined in a mastercurve of the form 
\begin{equation}
\eta(T,\dot{\gamma}) = \eta_0(T) {\cal F} (\dot{\gamma} \tau_\alpha(T)),
\label{eq:master}
\end{equation}
where the scaling function obeys ${\cal F}(x \to 0) \sim 1$ and ${\cal F}(x \gg 1) \sim 1/x$, in order to recover the two limits discussed above. Such a functional form has been used in many different systems~\cite{larson1999structure,sollich1997rheology,berthier2000two,berthier2002nonequilibrium,yamamoto1998dynamics}.

The scaling form in Eq.~(\ref{eq:master}) can be used to estimate large relaxation times using a non-linear shear flow. This is illustrated by a recent computational study by Jadhao and Robbins~\cite{jadhao2017probing,jadhao2019rheological} who measured the viscosity of a numerical model for squalane over a dynamic range of about four orders of magnitude at various temperatures in both the Newtonian and non-Newtonian regimes. Constructing a mastercurve of the family described by Eq.~(\ref{eq:master}), they used strongly non-linear measurements to infer the Newtonian viscosity in a temperature regime where the linear regime is too difficult to access directly. As a result, they reconstructed the Newtonian viscosity of their model over a window of about 25 orders of magnitude. From these extrapolated viscosity measurements, they concluded that the viscosity of squalane undergoes a crossover from fragile to strong behaviour at very low temperatures. 

Experimentally, this method has not yet been used to estimate large relaxation times, to our knowledge. In practice for molecular liquids, it is not obvious that this can be practically accomplished, as shearing highly viscous glassy materials often leads to inhomogeneous flows, including shear bands and fracture.  
  
{\it Critical discussion.} Using an equation such as Eq.~(\ref{eq:master}) requires an extrapolation along an assumed scaling function. As usual, the quality of the scaling function can only be tested when both regimes of the scaling form can be accessed directly. Instead, at very low $T$, the non-linear measurements can only access the non-equilibrium branch of the mastercurve, and forcing the measured data on a mastercurve requires both a horizontal and vertical adjustment that are in fact not independent and determine the value of $\tau_\alpha(T)$ extracted by this procedure. There is therefore no guarantee about the quality of the extrapolation. Moreover, given that the external shear flow strongly perturbs both the static and dynamic properties of the system, it seems physically unlikely that a strongly sheared system contains information about the equilibrium structure and dynamics of an unsheared system that relaxes many orders of magnitude more slowly. In practice, only data that are ``not too far'' from the crossover region captured by Eq.~(\ref{eq:master}) can fully be trusted, but it seems difficult to formulate a more precise, quantitative criterion to decide which data to trust, and which data to discard. The procedure thus appears too hazardous to allow strong statements about the functional form of the temperature dependence of the viscosity (and the relaxation time) at very low temperatures. 

\subsection{Modification of the pair interaction}

\label{sec:witten}

Another strategy to measure the dynamics at a given state point is to modify the pair interaction between the particles, which is of course easily done in a computer simulation. 

Assuming that particle crowding is responsible for slow relaxation, it is clear that making the particles less ``rigid'' should speed up the dynamics.
If one fully understands the crossover between the original system and its softened version, one can try to extrapolate from the dynamics of the softened system to the dynamics of the original system. 

A specific example of such a strategy can be found in the numerical study of Refs.~\cite{berthier2009compressing,berthier2009glass}, which addresses the glass transition of hard and soft particles in a unified manner. 
Here the original pair interaction one wishes to analyse is the hard sphere potential, similar to the one described in Sec.~\ref{sec:model}. As described above, the relaxation time only depends on a single parameter, which can be either $Z$ or $\phi$, since both are related by the equilibrium equation of state, $Z=Z(\phi)$. 

In their study, Berthier and Witten introduce a softened version of the hard sphere potential, which is chosen as a harmonic repulsion, $V(r) = \epsilon (1 - r/\sigma)^2$ for $r\leq \sigma$, so that the hard sphere potential is recovered in the limit $\epsilon \to \infty$. When $\epsilon$ (or, rather, the ratio $\epsilon/ T$) is finite, however, temperature and density can be varied independently, as in ordinary liquids. 

A possible strategy to extend the dynamic regime accessible by direct studies of hard spheres is to perform equilibrium measurements of the relaxation time of the harmonic sphere system at a series of state points $(\phi, \epsilon/T)$. The relaxation time $\tau_\alpha(\phi, \epsilon/T)$ of harmonic spheres converges, by construction, to the hard sphere relaxation time $\tau_\alpha^{hs}(\phi)$ in the appropriate limit: 
\begin{equation}
\tau_\alpha^{hs} (\phi) = \lim_{\epsilon/T \to \infty} \tau_\alpha(\phi, \epsilon/T).  
\label{eq:hs}
\end{equation}
There is no assumption behind Eq.~(\ref{eq:hs}).
Now, assuming a mastercurve of the form 
\begin{equation}
\tau_\alpha(\phi,\epsilon/T) = \tau_\alpha^{hs}(\phi) {\cal F}(X(\phi,\epsilon/T)),
\label{eq:scalingmct}
\end{equation}
\rev{where $X(\phi,\epsilon/T)$ is a simple function of its two variables and ${\cal F}(X(\phi,\infty))=1$, it is possible to construct the mastercurve ${\cal F}(X)$ in a regime} where direct measurements are possible, and a data collapse at larger $\phi$ can be used to infer the hard sphere relaxation time in a regime where its direct measurement is no longer possible. First employed in computer simulations~\cite{berthier2009compressing,berthier2009glass,schmiedeberg2011mapping}, the crossover between harmonic and hard spheres in the context of glassy dynamics has subsequently been studied theoretically using the mode-coupling theory of the glass transition~\cite{berthier2010scaling}. The theoretical study confirmed that a scaling form such as Eq.~(\ref{eq:scalingmct}) can indeed be generically expected.  

{\it Critical discussion.} As with several methods discussed above, the extrapolation using Eq.~(\ref{eq:scalingmct}) relies on an accurate determination of the scaling function used to rescale the data. The more problematic part comes from extrapolating data along the mastercurve in a regime where only a small part of the curve is covered by the actual data, so that the extrapolation is no longer reliable. As mentioned above, it is difficult to decide when the extrapolation becomes too hazardous to be performed, and it is tempting to use it to get a very broad range of timescales, with a reliability that is however totally unknown and impossible to assess. 

\subsection{Increasing the temperature: Glass melting}

\label{sec:melting}

Given that the dynamic relaxation at the equilibrium temperature $T$ is too long to be measured, a natural idea to speed up the relaxation is simply to heat the system to some higher temperature, $T'>T$, and measure the relaxation dynamics there immediately after the temperature has been changed. If one waits too long after the quench before measuring the dynamics, then one simply measures the equilibrium relaxation time at the new temperature $T'$. We are interested in the regime where $\tau_\alpha(T)$ is so large that relaxation is too long, i.e. the system appears arrested at temperature $T$. To detect some dynamics, $T'$ necessarily needs to be large enough for relaxation to proceed on the experimental timescale, so that the system will eventually transform back into the equilibrium liquid state at temperature $T'$. We call this transformation process ``glass melting''. It has been extensively studied recently in the context of ultrastable glasses~\cite{kearns2010one,sepulveda2014role,rodriguez2016relaxation,fullerton2017density,vila2020nucleation}. 

The ideal protocol is then: (i) prepare the equilibrium system at $T$; (ii) suddenly heat the system at $T'>T$ at time $t_w=0$; (iii) measure the relaxation dynamics right after the temperature change using some time correlation $Q(t+t_w,t_w)$ which quantifies the dynamics between times $t_w$ and $t+t_w$. 

The decay of the correlation $Q(t+t_w,t_w)$ with $t$ for $t_w=0$ allows us to define the melting time, $t_{melt}(T',T)$, which quantifies how long it takes the system to transform from the original glass created at $T$, into the supercooled liquid at $T'$. The ratio ${\cal S} \equiv t_{melt}(T',T) / \tau_\alpha(T')$ is the stability ratio~\cite{sepulveda2014role}. This adimensional number quantifies the kinetic stability of the initial amorphous state. The most stable systems prepared to date, either by vapor deposition or by the swap algorithm exhibit ${\cal S} = 10^4 - 10^6$~\cite{sepulveda2014role,fullerton2017density}. To determine the large relaxation time of the initial amorphous state, the idea is to study the evolution of $t_{melt}(T',T)$ with $T'$, and to extrapolate to the limit $t_{melt} (T' \to T, T) = \tau_\alpha(T)$.

For the method to be useful, however, the temperature jump $T \to T'$ must be large enough to considerably speed up the dynamics. There is no reason, therefore, for the dynamics to proceed during the melting process as if the system were close to equilibrium. A recent combination of theoretical~\cite{wolynes2009spatiotemporal,jack2016melting,jack2016note}, numerical~\cite{fullerton2017density,gutierrez2016front}, and experimental~\cite{kearns2010one,vila2020nucleation} studies suggests that the melting of ultrastable glasses indeed proceeds via a nucleation and growth process, where melting is initiated at rare regions within the sample from which the liquid invades the rest of the sample, very much like a crystalline material would also melt. The melting process should thus be characterized by spatially heterogeneous dynamics associated to a length scale which is considerably larger than in the equilibrium relaxation process~\cite{kearns2010one}. Regarding the behaviour of $t_{melt}(T',T)$ itself, it should depend in a non-trivial way both on the nucleation rate of the initial fluid droplets and on the velocity of the liquid front propagating through the glass~\cite{wolynes2009spatiotemporal,jack2016melting}, and it is accordingly difficult to capture these dependences using simple mathematical expressions. Empirically, it makes sense to expect some kind of activated behaviour~\cite{rodriguez2016relaxation}, 
\begin{equation}
t_{melt}(T',T) \sim \tau_{melt}(T',T) \exp \left( \frac{E(T',T)}{T'} \right), 
\label{eq:melt}
\end{equation} 
where the time prefactor $\tau_{melt}$ and the activation energy $E$ can in principle depend on both $T'$ and $T$. 

\begin{figure}
\includegraphics[width=8.5cm]{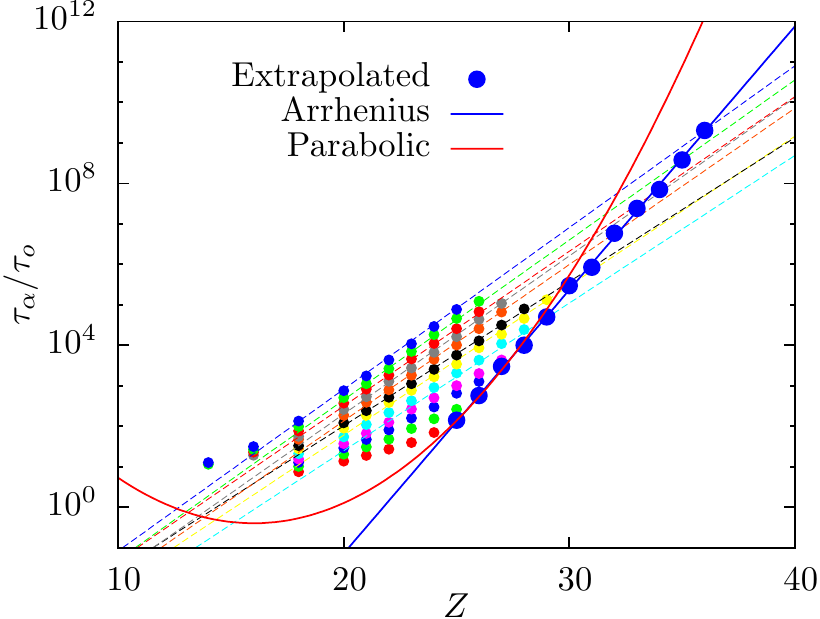}
\includegraphics[width=8.2cm]{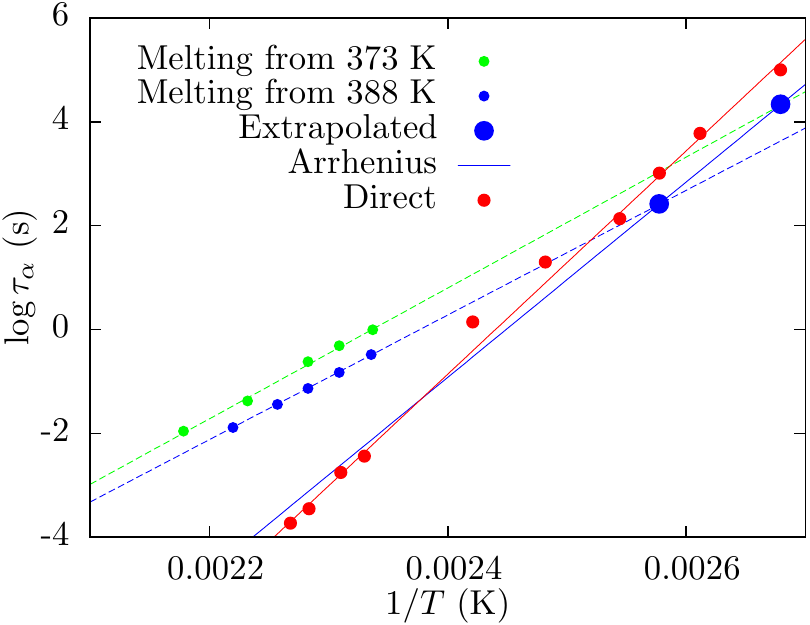}
\caption{Extrapolation of the glass melting times (small symbols) in hard sphere simulations (top) and experiments (bottom) on a metallic glass, using Eq.~(\ref{eq:melt2}) shown as dashed lines. \rev{In this figure, symbols with the same colour are obtained by changing $T'$ at constant $T$ with the colour code identical to Fig.~\ref{fig:mc}.}  
For the simulations, the extrapolated data (large blue symbols) fall below the parabolic fit (red line), and are described by an Arrhenius behaviour (blue line). Similar behavior is observed in experiments on a Au-based metallic glass~\cite{wang2016increasing}, with the extrapolated data falling below the direct measurements. }
\label{fig:melting}
\end{figure}

We have applied this method in the simulations of the $d=3$ hard sphere system.
Numerically, we first prepare equilibrium configurations at temperature \rev{$T$ (i.e. pressure $Z$) using the swap Monte Carlo algorithm, and we suddenly change it to $T' > T$ (i.e. $Z'<Z$). (Recall that for our system, the variables $Z$ and $1/T$ can be interchanged.) After the quench, we immediately record} the time evolution of the overlap function $Q(t+t_w,t_w)$ with $t_w=0$. From its decay time to the value $1/e$, we extract a melting time $t_{melt}(T',T)$ which we report in Fig.~\ref{fig:melting} for a series of different temperatures $T$. \rev{To construct this figure, we vary $T'$ for each value of $T$ and produce symbols with the same colour.} 
Empirically, we find that the behaviour of $t_{melt}$ for our system is well described by an Arrhenius behaviour, 
\begin{equation}
t_{melt}(T',T) \sim \tau_{melt}(T) \exp \left( \frac{E(T)}{T'} \right), 
\label{eq:melt2}
\end{equation} 
where the prefactor and the activation energy only depend on the stability of the initial configuration. Fits to Eq.~(\ref{eq:melt2}) are shown as dashed lines in Fig.~\ref{fig:melting}. We can then use these fitted Arrhenius functions to extrapolate the relaxation time up to temperature $T$, $\tau_\alpha (T) = t_{melt}(T' \to T,T)$. We report these extrapolated data in Fig.~\ref{fig:melting} as the larger blue symbols, along with the reference parabolic fit. We find that for the lowest temperatures $T$ where the extrapolation is the largest, the extrapolated data seem to follow an Arrhenius form quite distinct from the (fragile) parabolic fit. 

Experiments have been performed to record $t_{melt}(T',T)$ for a number of substances including both ultrastable and ordinary glasses~\cite{rodriguez2016relaxation,sepulveda2014role,wang2016increasing}. A fragile fitting function as in Eq.~(\ref{eq:melt}) has been used to fit the data~\cite{rodriguez2016relaxation}. The experimental melting times are qualitatively consistent with the simulation results reported in Fig.~\ref{fig:melting} in that an Arrhenius extrapolation of melting times from high temperature yields estimate equilibrium relaxation times that are too short.  Simple, classic models of glassy dynamics~\cite{diezemann2011memory,tool1931variations,narayanaswamy1971model,moynihan1976structural} also capture this behavior. 

One set of experimental results is particularly useful to compare with our simulation results. Fast-scanning calorimetry was used to study the melting of a Au-based metallic glass~\rev{\cite{wang2016increasing}.} Samples were first taken to equilibrium and then very quickly heated to the annealing temperatures, where isothermal melting was observed. These experimental results are included in the second panel of Fig.~\ref{fig:melting}, plotted in the same format as the simulation results. In these experiments, the Arrhenius extrapolation procedure (of about 2 orders of magnitude) yields an estimate of $\tau_\alpha$ about 3 times shorter than the measured value of $\tau_\alpha$. 

{\it Critical discussion.} When $T'$ and $T$ are very different, the physical processes governing the glass melting and the equilibrium relaxation are distinct. In particular, glass melting is dominated at short times by nucleation events at rare locations that relax much faster than the rest of the system. It is unclear how these fast events contribute to the structural relaxation time in equilibrium conditions. In equilibrium, fast localised processes are often considered to simply contribute to secondary processes~\cite{yu2017structural}. For glass melting, the extrapolation of the temperature dependence of these processes is unlikely to provide the correct estimate of the equilibrium relaxation time $\tau_\alpha$, as indeed both numerical and experimental data in Fig.~\ref{fig:melting} suggest. Presumably, when $T'$ and $T$ are close enough, the Arrhenius behaviour in Eq.~(\ref{eq:melt}) will break down and increase faster to provide the right limit for $t_{melt}(T' \to T,T)$~\cite{rodriguez2016relaxation}. 

\subsection{Lowering the energy barriers}

\label{sec:epsilon}

We now discuss an alternative method to speed up the dynamics that is not directly applicable to experiment, but can be implemented numerically relatively easily. It has recently been used to study the dynamics of spin glasses in computer simulations~\cite{pemartin2019numerical}.

The method \rev{directly} attacks the problem we want to solve: starting from an equilibrium configuration at a given temperature $T$, thermal fluctuations are insufficient to trigger structural relaxation in the available time window, $t_{exp}$. An obvious consequence is that only configurations which are close to the initial condition in configuration space are then sampled by the dynamics. The idea here is therefore to add a thermodynamic field, denoted $\epsilon$, which energetically disfavors configurations that are too close to the initial condition. This is equivalent to increasing the free energy of the metabasin the system was visiting at time $t=0$ to induce a faster relaxation towards another metabasin with a lower free energy.

\begin{figure}
\includegraphics[width=8.5cm]{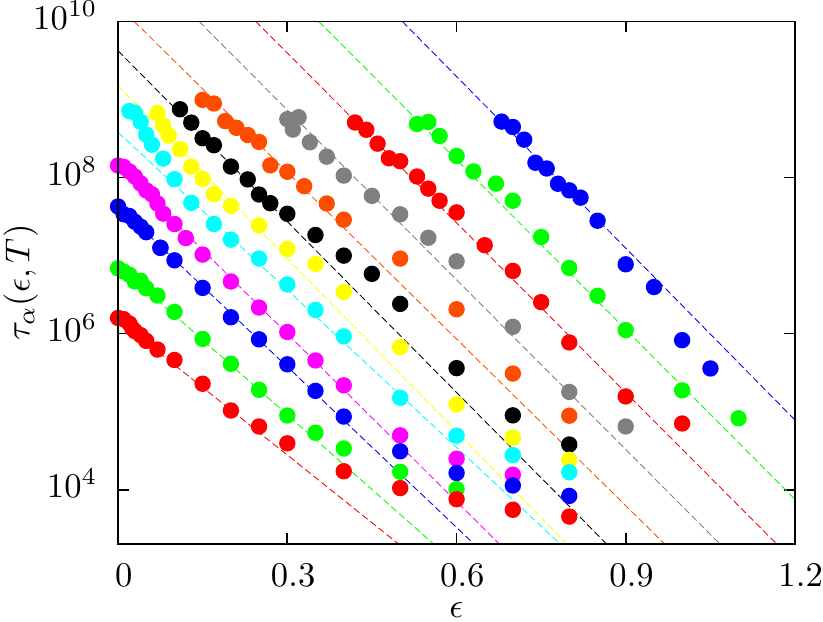}
\includegraphics[width=8.5cm]{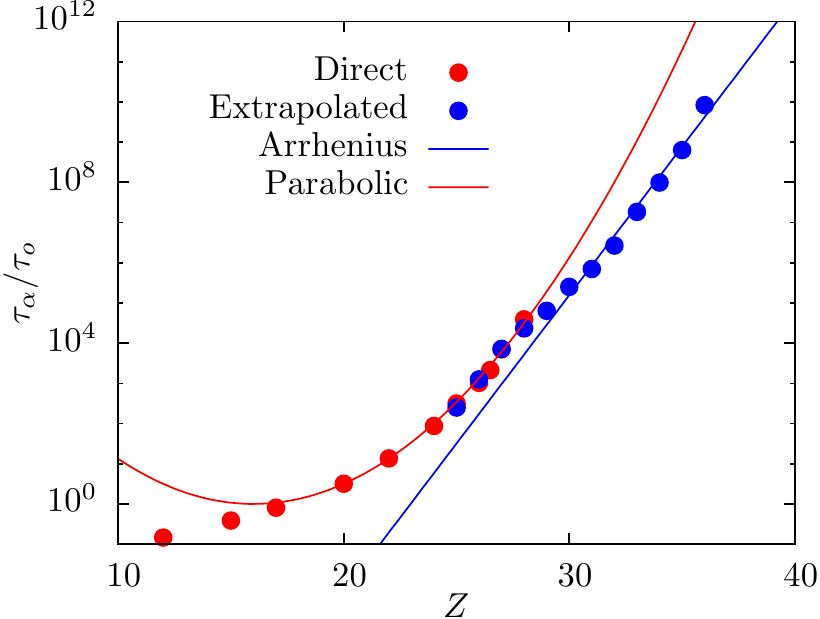}
\caption{Top: Evolution of the relaxation time $\tau_\alpha(\epsilon,T)$ with the amplitude of the field $\epsilon$ at constant temperature (colour code as in Fig.~\ref{fig:melting}). Bottom: The extrapolated times $\tau_\alpha(\epsilon \to 0,T)$ are reported together with the direct dynamical measurements and the reference parabolic fit, and are described by an Arrhenius behaviour (blue line).}
\label{fig:epsilon}
\end{figure}

In practice, we add a thermodynamic perturbation to the model of the form
\begin{equation}
\Delta E = \epsilon Q(t,0), 
\end{equation}
where $Q(t,0)$ is the dynamic overlap defined in Eq.~(\ref{eq:overlap}) between a configuration at time $t$ and the initial condition at time $t=0$. 
The field $\epsilon>0$ is such that large $Q(t,0)$ values (which correspond to configurations close to the initial condition) cost an additional energy $\Delta E \sim \epsilon$, whereas low $Q(t,0)$ values (which correspond to configurations far from the initial configuration) are not affected by the field $\epsilon$. 

Therefore the field $\epsilon$ acts as a thermodynamic force that repels the system from its initial condition, forcing it to relax. The idea is to measure the relaxation time in the presence of the field at temperature $T$, $\tau_\alpha(\epsilon,T)$, and to then extrapolate its behaviour towards the equilibrium system at $\epsilon \to 0$. Conceptually, this approach is not very different from the temperature change in the previous section, and here again large field values will be needed to speed up the dynamics in the interesting low-temperature region, which very likely will drive the dynamics far from the equilibrium relaxation process, making the extrapolation back to equilibrium once again potentially problematic. 

The method is illustrated in Fig.~\ref{fig:epsilon} for hard spheres. The top panel shows how $\tau(\epsilon,T)$ becomes smaller as $\epsilon$ increases, demonstrating that indeed the relaxation from the initial condition is considerably accelerated by the field $\epsilon$ (notice that the vertical time axis is in log-scale). To extrapolate the data we use a simple Arrhenius form
\begin{equation}
\tau_\alpha(\epsilon,T) \sim \tau_\alpha(T) \exp \left( - \frac{\epsilon}{E_0(T)} \right),
\label{eq:epsilon}
\end{equation}
where the prefactor gives the desired equilibrium relaxation time, and $E_0(T)$ is a fitted energy scale. The fit is excellent both at high and low temperatures, but interesting deviations can be seen in the crossover regime between the two, especially near $\epsilon \to 0$.   

The extrapolated relaxation times are shown in Fig.~\ref{fig:epsilon} together with the direct measurements and the parabolic reference fit. We observe, similar to the melting times in Sec.~\ref{sec:melting}, that the extrapolated timescales are better described at low temperatures by a simple Arrhenius dependence. The deviations from the Arrhenius fit of Eq.~(\ref{eq:epsilon}) observed for intermediate temperature indicate that the extrapolation performed at lower temperatures considerably underestimates the true relaxation time. Physically, this presumably results from the fact that Eq.~(\ref{eq:epsilon})  only describes the strongly non-equilibrium relaxation at large $\epsilon$, when the dynamics differs qualitatively from the equilibrium one. 

{\it Critical discussion.} The same discussion as for the glass melting is relevant here, in the sense that a strong $\epsilon$ field is needed to speed up the dynamics at low $T$, but such strong field drives the system far from equilibrium, which renders the extrapolation back to equilibrium problematic. As for the temperature change, it is likely that the field $\epsilon$ induces dynamics where rare soft regions first relax and then trigger the relaxation elsewhere in the system, which could explain the Arrhenius dependence of the extrapolated relaxation times in Fig.~\ref{fig:epsilon}.

\subsection{Aging dynamics}

\label{sec:aging}

Glasses equilibrated at very low temperatures barely relax when left at that temperature, which is our central problem. It is therefore common practice to change the temperature and study the subsequent dynamics. The glass melting process discussed above in Sec.~\ref{sec:melting} is a specific example of such a protocol in which the dynamics is followed all the way back to equilibrium, in order to infer a melting time. Quite often, however, aging experiments do not reach complete equilibrium, and the time dependence is carefully studied to infer information about the equilibrium situation~\cite{struik1977physical}. This is the topic of the present section. We distinguish between strategies where some intermediate time quantity is followed, and those which directly attempt to follow the evolution of the structural relaxation time.  

\subsubsection{Intermediate time dynamics}

Ediger and coworkers~\cite{kasting2019relationship,yu2015suppression} have followed the evolution of the dynamics in the frequency range corresponding to the $\beta$ process in several molecular glass-formers to infer the relaxation time of ultrastable glasses by comparing two distinct sets of experiments, based on dielectric measurements, $\chi''(\omega)$: (i) perform an isothermal aging experiment after a rapid quench from high temperature to a final temperature $T$, and measure how the amplitude of the $\beta$ process evolves with the waiting time $t_w$ spent since the quench; (ii) directly measure the amplitude of the $\beta$ process for the equilibrium ultrastable glass at the same temperature $T$. By construction, the data in the first set of measurements must (slowly) converge to the result of the second set in the limit of large $t_w$. Estimating how long it takes to the first set of measurements to match the second one provides the equilibration time at temperature $T$. In practice, Kasting {\it et al.} observe that $\chi'' \sim - \log t_w$ during the aging, and they extrapolate (in some cases by more than 6 orders of magnitude) this logarithmic behaviour to the independently measured equilibrium value. The extrapolated relaxation times deviate significantly from a fit performed for $T>T_g$ using the VFT law~\cite{kasting2019relationship}.

We have followed a similar strategy in hard sphere simulations to infer long relaxation times using intermediate times dynamics. To this end, we follow the spirit of Ref.~\cite{kasting2019relationship} and compare two independent sets of numerical measurements. First, we measure the dynamics of equilibrium samples at various pressures, $Z$, over a large time window. We obtain the average correlator $Q(t)$ in equilibrium. The analog of the out-of-phase component of the dielectric susceptibility is computed as~\cite{berthier2005numerical} 
\begin{equation}
Q''(\omega) =  \int_{-\infty}^\infty d t  
\frac{dQ(t)}{dt} \frac{\omega t}{1+ (\omega t)^2}.
\end{equation}
The spectrum $Q''(\omega)$ exhibits the usual two-peak structure, corresponding to the structural relaxation at low frequency and to the microscopic relaxation process at large frequencies. The two peaks are separated by a minimum at intermediate frequency. We denote the amplitude of the minimum as $Q''_{min}$, and take this as our intermediate-time dynamical quantity. At equilibrium, $Q''_{min,eq}(T)$ only depends on temperature. Our choice of observable is justified by the fact that it avoids choosing an arbitrary frequency scale to probe the intermediate time dynamics. 

\begin{figure}
\includegraphics[width=8.5cm]{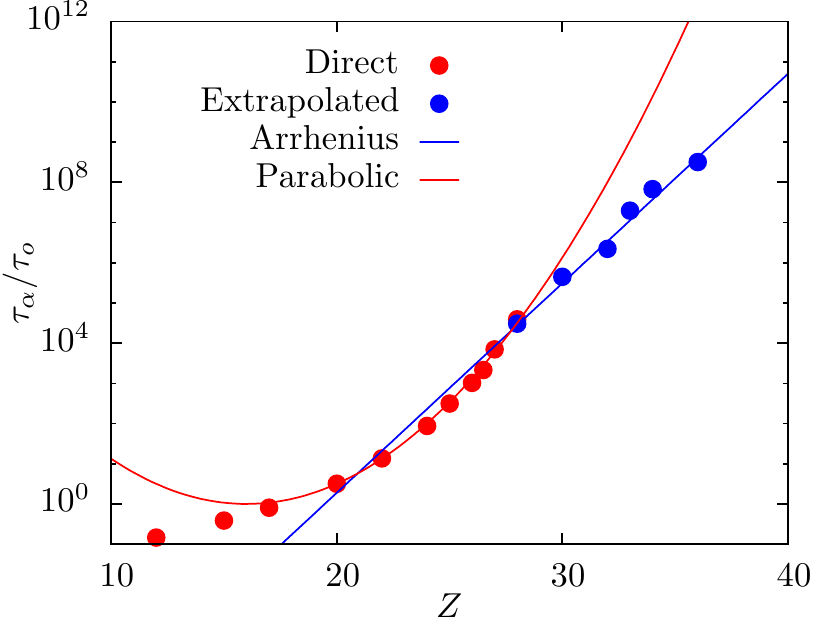}
\caption{Estimates of the relaxation time using the aging of the intermediate time dynamics. The extrapolated times are reported together with the direct dynamical measurements and the reference parabolic fit, and are
described by an Arrhenius behaviour (blue line).}
\label{fig:beta}
\end{figure}

In a second set of measurements, we perform an instantaneous quench from a high temperature ($Z=17$; recall from Fig.~\ref{fig:bulk} that the onset is near $Z=20$) to a series of low temperatures, and follow the isothermal evolution of $Q''_{min}(T,t_w)$ with the time $t_w$ spent since the quench. Empirically, we find that $Q''_{min}(T,t_w) / Q''_{min,eq}(T) \approx A (T) t_w^{-\alpha}$, with $\alpha \approx 0.23$ and $A(T)$ a temperature dependent prefactor. We then use this functional form to estimate the time when the dynamics comes to equilibrium, $Q''_{min}(T,t_w=\tau_\alpha) / Q''_{min,eq}(T) = 1$, and use this as an extrapolated estimate for the equilibrium relaxation time of the system. We report the results of this analysis in Fig.~\ref{fig:beta}, where the maximum extrapolated relaxation time is about $10^5$ times larger than the maximal waiting time $t_w$ accessible in the simulations. 

As expected, the equilibration time defined in this manner matches the relaxation time data in the regime where $\tau_\alpha$ can be measured directly. However, when extrapolated to lower temperatures, the estimated relaxation times seem to grow in an Arrhenius manner, with an activation energy that is significantly smaller than the one obtained by fitting in Sec.~\ref{sec:temperature}.  

{\it Critical discussion.} 
These measurements correspond to a non-equilibrium version of the ones described in Sec.~\ref{sec:beta} for equilibrium situations, but instead of using a relation between $\tau_\beta$ and $\tau_\alpha$, one simply uses the dynamics at intermediate timescale as an observable which should reach its equilibrium behaviour for aging times $t_w \sim \tau_\alpha$. Thus, no particular relationship between the $\alpha$ and $\beta$ relaxations is needed. However, there are two key assumptions. One is that equilibration of the $\beta$ process is achieved on a timescale governed by $\tau_\alpha$, which appears validated by experiments and simulations, at least in the regime where this can be directly tested. The second key assumption is that observing the dynamics over a given window of waiting times $t_w$ allows one to extrapolate $\tau_\alpha$, even when the ratio $\tau_\alpha /t_w \gg 1$. The simulations above indicate that the extrapolated relaxation times are underestimated by the approach. This makes physical sense since at early aging times, the system has no reason to explore the same relaxation processes (or, equivalently, the same part of configuration space) that govern the dynamics at much later timescales. 

\subsubsection{Long-time dynamics}

We now focus on aging experiments which use protocols similar to the previous section, but directly focus on the long-time dynamics rather than on an intermediate time quantity. 

This type of analysis will thus combine several procedures mentioned above. (i) Dynamics is followed after a sudden change in the experimental conditions, such as a temperature change, as in Sec.~\ref{sec:melting}; (ii) the dynamics is measured over the largest available time window, as in Sec.~\ref{sec:TTS}; (iii) a rescaling of the obtained dynamics on some mastercurve is performed, as in Secs.~\ref{sec:shear} and \ref{sec:witten}. Therefore, the hypotheses needed in this case are a combination of all those listed in the corresponding sections.  

This type of analysis was recently performed on both 20-million-year old amber glasses~\cite{zhao2013using} and vapor deposited ultrastable teflon~\cite{yoon2018testing}. Earlier, extremely long aging experiments have also been performed on liquid-cooled glasses~\cite{zhao2012temperature,richert2013comment,zhao2013response} and analysed along similar ideas. In the teflon work, mechanical measurements performed over a large timescale of about $10^4$~s are used to infer equilibrium relaxation times up to $10^{15}$~s, i.e. an extrapolation of 11 orders of magnitude. 
    
Using hard sphere simulations, we performed an analysis of the aging dynamics. The thermal history is an ordinary, instantaneous quench from high temperature ($Z=17$) to some low temperature. During the aging, the volume of the system increases slowly towards its equilibrium value. For very low temperatures, the relaxation time is very large and the volume does not reach a steady state during the aging, and depends explicitly on $t_w$, the time spent since the quench. We have recorded the evolution of the volume fraction $\phi(t_w)$ (which scales as the inverse of the volume). In a separate set of simulations, we use the swap Monte Carlo algorithm to directly measure $\phi_{eq}$ in equilibrium. We can then define the adimensional variable $\phi_{eq} - \phi(t_w)$, which must vanish, by construction, when $t_w \gg \tau_\alpha$. We use this variable to infer a relaxation time at low temperature assuming a mastercurve of the form, 
\begin{equation}
\phi_{eq} - \phi(t_w) = {\cal G}(t_w / \tau_\alpha),
\label{eq:scalingaging}
\end{equation}     
where ${\cal G}(x)$ is a scaling function such that ${\cal G}(x \to \infty)=0$, to reflect equilibrium.  

The knowledge of the equilibrium value of the volume fraction (or, of any alternative choice of a physical observable) is often not available experimentally, and the rescaling onto mastercurves is thus less constrained. In practice, experimental data collapse can be achieved by shifting the measurements both horizontally and vertically to produce a mastercurve, see for instance Ref.~\cite{zhao2013using}. 

\begin{figure}
\includegraphics[height=3.96cm]{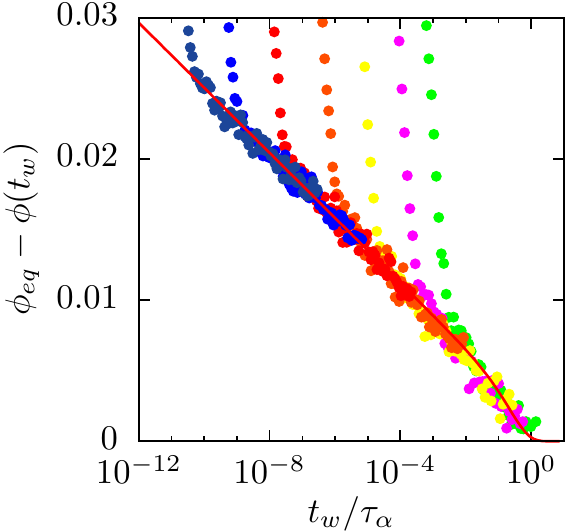}
\includegraphics[height=3.88cm]{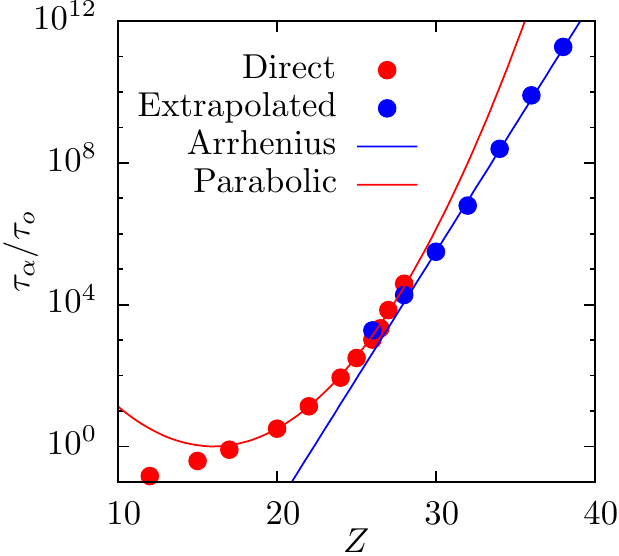}
\caption{Left: Evolution of the volume fraction with waiting time after a quench from high temperature ($Z=17$) to various final states \rev{(shown with different colors, the final pressure increases from right to left from $Z=26$ to $Z=38$)} in hard sphere simulations. The waiting time is rescaled to provide the best data collapse, assuming that Eq.~(\ref{eq:scalingaging}) is valid. The full line illustrates the scaling function ${\cal G}(x)$.
Right: The extrapolated times $\tau_\alpha$ are described by an Arrhenius behaviour (blue line) and are reported together with the direct dynamical measurements and the reference parabolic fit.}
\label{fig:aging}
\end{figure}
 
The numerical analysis is summarised in Fig.~\ref{fig:aging}, where $\tau_\alpha$ is adjusted for each volume fraction to get the best data collapse, in the spirit of Refs.~\cite{zhao2013using,yoon2018testing}. The data seem to follow a logarithmic dependence at short times, ${\cal G}(x\ll1) \sim - \log(x)$, interrupted near $x=1$ by a (possibly stretched) exponential cutoff, as illustrated in Fig.~\ref{fig:aging}. Since the procedure only provides $\tau_\alpha$ up to a global rescaling factor, we include in the analysis one temperature for which the relaxation time can be measured directly in equilibrium to set the overall scale, as is also done experimentally.

As in the experiments (and similarly to Secs.~\ref{sec:shear}, \ref{sec:witten}), the dynamics is actually measured over a fixed waiting time window, but large relaxation times are inferred by assuming the rescaling form in Eq.~(\ref{eq:scalingaging}). The scaled variable $t_w/\tau_\alpha$ covers about 12 orders of magnitude in Fig.~\ref{fig:aging}, even though the data for any single temperature covers about 4 orders of magnitude. The mastercurve is thus artificially reconstructed by gluing together small pieces of data. 

We finally report $\tau_\alpha$ obtained through this analysis in Fig.~\ref{fig:aging}, along with the direct measurements and the reference parabolic fits. The extrapolated data deviate rapidly from the parabolic fit and follow a simpler Arrhenius form. We note that the temperature dependence that we extract from this analysis is Arrhenius and, in this respect, similar to the results obtained in several experimental studies~\cite{zhao2012temperature,zhao2013using,yoon2018testing}, even though in our system the dynamics in the extrapolated regime is expected to be slower than indicated by this Arrhenius law.   

{\it Critical discussion,} The method presented here combines many of the tools discussed separately in earlier sections, and thus the criticisms described there all apply simultaneously here. First, this is a non-equilibrium method, and there is no guarantee that the dynamics is the same as in equilibrium. Second, the measured time window is short compared to the long-time dynamics inferred using a reconstructed mastercurve of unknown validity. Third, it is possible that these dynamics are influenced by the existence of secondary processes which have a temperature dependence weaker than structural relaxation. 

\section{Conclusion}

\label{sec:conclusion} 

Measuring the relaxation time of very stable or very old glassy materials is currently an important challenge both for simulations and experiments~\cite{mckenna2015accumulating,ediger2017perspective,mckenna2020glass}. The title of this article is intentionally provocative, since it is by definition impossible to measure a relaxation time that is too long to be measured. No method can safely estimate very large timescales, and no method can deliver rigorous lower or upper bounds for $\tau_\alpha$. All such measurements need to be critically analysed, and all physical conclusions subsequently drawn from these extrapolations need to be critically evaluated.

\begin{figure}
\includegraphics[width=8.5cm]{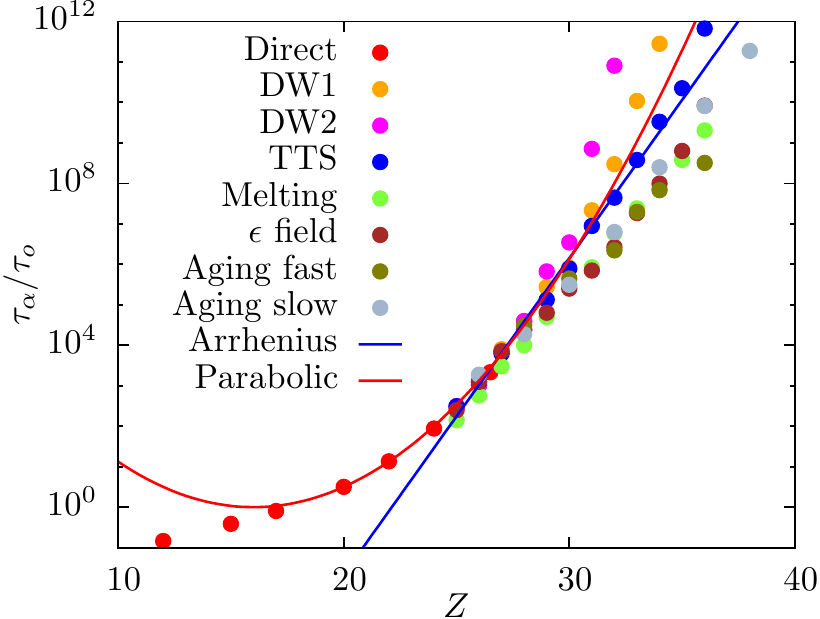}
\caption{Compiliation of extrapolated $\tau_\alpha$ values obtained by the methods described in this article. The parabolic fit represents our reference temperature dependence, while the Arrhenius fit obtained in Sec.~\ref{sec:temperature} should be considered a lower bound. Most non-equilibrium extrapolations fall below the Arrhenius fit, and are likely in error.} 
\label{fig:all}
\end{figure}
 
In Fig.~\ref{fig:all}, we gather all the methods used throughout the paper to extrapolate the relaxation time of the hard sphere system from $\tau_\alpha/\tau_o = 10^5$ (end of numerical time window) to $\tau_\alpha/\tau_o = 10^{12}$ (experimental glass transition temperature). Starting from the temperature fit extrapolation, we suggested that the parabolic fit could be used as a reference for later comparisons, whereas the Arrhenius fit should be seen as a reasonable (but non-rigorous) lower bound since it neglects the curvature of the actual data. The most obvious statement regarding Fig.~\ref{fig:all} is that the relaxation times estimated by these methods differ substantially and, at best, only one of these methods can be correct.

We find that the extrapolations using the Debye Waller factor (Sec.~\ref{sec:DW}), and time temperature superposition at equilibrium over a large time window (Sec.~\ref{sec:TTS}) provide extrapolations above the Arrhenius fit and have some degree of fragility, comparable to the parabolic fit extrapolation. 

On the other hand, several non-linear and non-equilibrium methods, such as glass melting (Sec.~\ref{sec:melting}), the $\epsilon$-field (Sec.~\ref{sec:epsilon}), aging dynamics (Sec.~\ref{sec:aging}) seem to lead to a strong underestimate of the actual relaxation time. As explained in the corresponding subsections, a strong underestimation of the extrapolated relaxation times makes physical sense, since these methods systematically probe physical processes that are distinct from the long-time equilibrium dynamics that they seek to estimate. \rev{This conclusion is reinforced by the fact that these non-equilibrium methods yield estimates that are smaller than the Arrhenius extrapolation of the relaxation times, which represents a well-motivated lower bound to the relaxation times.} 

\rev{Our general conclusion is that extrapolating to obtain large relaxation times is a difficult exercise which seems to be giving consistent results only if the extrapolation involves a modest number of decades, but becomes a hazardous task when a larger number of decades is involved. Given all the uncertainties mentioned throughout this article, it should be clear that one should not use extrapolated data to make definitive statements about the fate of supercooled liquids at very low temperatures, such as addressing for instance the existence of a Kauzmann transition.}

\rev{We would like to offer some more optimistic views before closing. First, all estimates (and presumably underestimates) of the relaxation times in vapor-deposited ultrastable glasses and in silico glasses prepared by the swap Monte Carlo algorithm indicate that these systems are characterised by relaxation times that are many orders of magnitude larger than ordinary glasses. Therefore, there should be no doubt that these materials do behave as extremely old glasses, that have been prepared in a comparatively modest amount of time. Second, even if we do not yet know how to accurately estimate the extremely long relaxation times of ultrastable glasses, study of these materials provides important clues.  For example, the density~\cite{beasley2019vapor}, modulus~\cite{fakhraai2011structural}, enthalpy~\cite{ramos2011character}, vibrational density of states~\cite{wang2019low}, and configurational entropy~\cite{berthier2017configurational,berthier2019zero} of ultrastable glasses are consistent with extrapolation from the supercooled liquid above $T_g$. This indirectly indicates that the relaxation time should similarly be increasing smoothly as the temperature decreases, consistent with all the extrapolations presented here.}

At present we do not see how to improve on the methods discussed here to estimate experimentally very large relaxation times. Regarding simulations, there is still the hope that novel algorithms can be invented to considerably speedup not only the equilibration of the system, but also its physical relaxation dynamics. Studying the dynamics at very long times is thus the very clear, next challenge faced by computer studies of glassy materials. 

\begin{acknowledgments}
We thank G. Parisi for suggesting the method proposed in Ref.~\cite{pemartin2019numerical}. We also thank B. Guiselin and C. Scalliet for several discussions and technical help with the simulations. Discussions with R. Jack, G. McKenna and G. Tarjus are also acknowledged. This work was supported by a grant from the Simons Foundation (Grant No. 454933, L. B.) and by NSF through the University of Wisconsin Materials Research Science and Engineering Center (Grant DMR-1720415, M. D. E.).
\end{acknowledgments}

\section*{Data availability}

The data that support the findings of this study are available from the corresponding author upon reasonable request.

\bibliography{glasses.bib}{}

\end{document}